\documentclass[traditabstract]{aa}

\usepackage{bm}
\usepackage{amsmath}
\usepackage{amssymb}
\usepackage{graphicx}
\usepackage{txfonts}
\usepackage{dcolumn}
\usepackage{natbib}
\usepackage{url}
\usepackage{enumitem}
\usepackage{bm}

\usepackage[usenames,dvipsnames]{color}

\bibpunct{(}{)}{;}{a}{}{,}

\newcommand{\diff}{{\mrm{d}}}
\newcommand{\Diff}{{\mrm{D}}}
\newcommand{\DD}[2]{\frac{\Diff #1}{\Diff #2}}
\newcommand{\pp}[2]{\frac{\partial #1}{\partial #2}}
\newcommand{\ppp}[2]{\frac{\partial^2 #1}{\partial #2}}

\newcommand{\mrm}[1]{{\mathrm{#1}}}

\begin{document}

 \title{Overshooting by differential heating}

 \author{R. Andr\'assy, H. C. Spruit}

 \institute{Max Planck Institute for Astrophysics, Karl-Schwarzschildstr. 1, 85748 Garching, Germany}

 \date{Received ; accepted }

 \abstract{
On the long nuclear time scale of stellar main-sequence evolution, even weak mixing processes can become relevant for redistributing chemical species in a star. We investigate a process of ``differential heating,'' which occurs when a temperature fluctuation propagates by radiative diffusion from the boundary of a convection zone into the adjacent radiative zone. The resulting perturbation of the hydrostatic equilibrium causes a flow that extends some distance from the convection zone. We study a simplified differential-heating problem with a static temperature fluctuation imposed on a solid boundary. The astrophysically relevant limit of a high Reynolds number and a low P\'eclet number (high thermal diffusivity) turns out to be interestingly non-intuitive. We derive a set of scaling relations for  the stationary differential heating flow. A numerical method adapted to a high dynamic range in flow amplitude needed to detect weak flows is presented. Our two-dimensional simulations show that the flow reaches a stationary state and confirm the analytic scaling relations. These imply that the flow speed drops abruptly to a negligible value at a finite height above the source of heating. We approximate the mixing rate due to the differential heating flow in a star by a height-dependent diffusion coefficient and show that this mixing extends about $4\%$ of the pressure scale height above the convective core of a $10\,M_\odot$ zero-age main sequence star.
 }

 \maketitle

\section{Introduction}

Our lack of understanding of (magneto)hydrodynamic transport processes in stars has hampered progress in developing the stellar evolution theory since its earliest beginnings. One particular aspect of the problem is the mixing in the boundary layers between convection and radiative zones in stellar interiors, which is also known as the problem of convective overshooting. Despite the indisputable advance in numerical simulations, the problem remains extremely challenging owing to the extreme range of the length and time scales involved in it.

The set of physical mechanisms that provide mixing at a convective/stable interface very likely depends on the type of convection zone involved. Because it is exposed to outer space at the top, a convective envelope is driven by the cold plumes originating in the photosphere. It is quite possible that the plumes span the whole convection zone and even provide mixing at its bottom boundary \citep[cf.][and references therein]{AndrassySpruit13}. A convective core or shell is, on the other hand, fully embedded in the star, its stratification is much weaker, and the temperature fluctuations within it are much smaller. Consequently, a different set of physical mechanisms may dominate mixing at its boundary. 

It has long been known that the kinetic energy of the low-Mach-number flow in a convective core (or shell) is so low that the convective motions are stopped within about one per cent of the pressure scale height once they enter the steep entropy gradient of the radiative zone \citep{Roxburgh65, SaslawSchwarzschild65}. The motions can reach much farther, though, if they are vigorous enough to flatten the radiative entropy gradient above the convective core. In this case, known as the process convective penetration, the motions gradually ``erode'' the radiative stratification on the thermal time scale until radiative diffusion stops any further advance of the erosion front \citep{ShavivSalpeter73, vanBallegooijen82, Zahn91}. Finally, the fluid parcels hitting the stable stratification always generates a spectrum of internal gravity waves, which may also provide a certain amount of mixing \citep{Press81, Garcia-LopezSpruit91, Schatzman96}.

Several of the processes mentioned may operate at the convection zone's boundary at the same time. Their effects on long time scales and at long distances from the boundary are very different. In full numerical hydrodynamic simulations, the restrictions on time scales that can be covered makes it difficult to disentangle these effects. Physical insight developed by different means is needed to extrapolate them to longer time scales and distances.

We take a closer look at one specific process operating at a convective/stable interface in the interior of a star. Thermal diffusion causes temperature fluctuations from the convection zone's boundary to spread into the stable stratification. Temperature differences on surfaces of constant pressure set up a flow even in the absence of momentum transport by hydrodynamic stress. We call this process ``differential heating'', explore the physics of it in an idealised set-up, and estimate what amount of mixing it could cause in the stellar interior.

\section{The differential heating problem}
\label{sec:differential_heating_problem}

\subsection{Problem formulation and simplification}
\label{sec:problem_formulation}

Consider a horizontal, solid surface with a stably-stratified fluid overlying it.\footnote{Equivalently, the stably stratified fluid could be placed under the differentially heated surface. The role of hot and cold spots on the surface would be reversed in this case. We discuss only one case for the sake of concreteness.} A temperature fluctuation imposed at the surface propagates into the fluid by a diffusive process and upsets the hydrostatic equilibrium. We investigate what the properties of the resulting flow are.

By replacing the convective/stable interface by a solid wall, we eliminate all the phenomena related to the inertia of the convective flows and the shear induced by them. This allows us to study the physics of differential heating in isolation. The upper boundary is taken far enough not to influence the flow. Next we introduce further assumptions to facilitate the mathematical description and the subsequent analysis of the problem:
\begin{enumerate}[label={(\arabic*)}]
\item The flow is confined to a layer that is significantly thinner than the pressure scale height.
\item The fluid is a chemically homogeneous, ideal gas.
\item The Brunt-V\"{a}is\"{a}l\"{a} (buoyancy) frequency of the stratification is constant.
\item Thermal diffusivity is constant.
\item The gravitational field is homogeneous.
\item The differentially heated surface is flat and horizontal.
\item The flow is constrained to two spatial dimensions.
\end{enumerate}

Assumption (1) allows us to use the Bussinesq approximation and turns out to be justified. The chemical homogeneity that we assume in (2) is, at least for the nuclear-burning layers in a star, only realistic at the onset of the burning. The differential heating process above a convective core would weaken as the nuclear burning progresses owing to the increase in the mean molecular weight in the core. We focus on the chemically homogeneous case to keep the number of parameters tractable. We introduce (3) and (4) for the same reason. The Brunt-V\"{a}is\"{a}l\"{a} frequency depends on the distance from the convective/stable interface in a real star. The constant frequency in our analysis can be thought of as a typical value for the layer influenced by differential heating. Finally, we add the last three assumptions to make our analysis more transparent and to reduce the computational costs of the numerical solutions. We have to keep in mind, however, that the constraint (7) might influence the stability properties of the flow, and thus some of our conclusions may not apply to the three-dimensional case.

The Bussinesq equations are \citep{SpiegelVeronis60}
\begin{align}
\bm\nabla \cdot \bm u &= 0, \label{eq:continuity1} \\
\DD{\bm u}{t} &= - \frac{1}{\rho_\mrm{m}}\bm\nabla p^\prime + \frac{T^\prime}{T_\mrm{m}} g \bm k + \nu\,\nabla^2 \bm u, \label{eq:momentum1} \\
\DD{T^\prime}{t} &= -\frac{T_\mrm{m} N^2}{g}w + \varkappa\,\nabla^2 T^\prime, \label{eq:energy1}
\end{align}
where $\bm u$ is the fluid velocity, $\Diff/\Diff t = \partial/\partial t + \bm u \cdot \bm\nabla$ the Lagrangian time derivative, $\rho_\mrm{m}$ and $T_\mrm{m}$ are the mean density and temperature, respectively, $p^\prime$ and $T^\prime$ the pressure and temperature perturbations, respectively, $g$ is the gravitational acceleration, $\bm k$ a unit vector pointed in the vertical direction, $\nu$ the kinematic viscosity, $N$ the Brunt-V\"{a}is\"{a}l\"{a} frequency, $w$ the vertical velocity component, and $\varkappa$ the thermal diffusivity.

Equations \ref{eq:continuity1}, \ref{eq:momentum1}, and \ref{eq:energy1} still contain several dimensional parameters. It is crucial to realise that there is a natural system of units for the differential heating problem that makes the equations dimensionless. The flow in this problem is set off by thermal diffusion in a stably stratified medium, therefore the inverse of the Brunt-V\"{a}is\"{a}l\"{a} frequency, $1/N$ (or a multiple of it), is a natural unit of time. Having made this choice, we can define a natural unit of distance as $\sqrt{\varkappa/N}$, which is a typical thermal-diffusion length scale on the time scale $1/N$. The dimensionless Bussinesq equations are then
\begin{align}
\bm\nabla \cdot \bm u &= 0, \label{eq:continuity2} \\
\DD{\bm u}{t} &= - \bm\nabla p + \vartheta\bm k + \Pr\,\nabla^2 \bm u, \label{eq:momentum2} \\
\DD{\vartheta}{t} &= -w + \nabla^2 \vartheta, \label{eq:energy2}
\end{align}
where we omit any symbol to indicate the new units. We have also introduced a new pressure-like variable $p = p^\prime/\rho_\mrm{m}$ and the buoyancy acceleration $\vartheta = g T^\prime/T_\mrm{m}$, which we continue to call the ``temperature fluctuation'' in the rest of the paper, because that is the central concept in the differential heating process. The Prandtl number \mbox{$\Pr = \nu/\varkappa$} now becomes a measure of kinematic viscosity, because the new unit of diffusivity is $\varkappa$. Equations~\ref{eq:continuity2}, \ref{eq:momentum2}, and \ref{eq:energy2} are particularly well suited to theoretical studies since their solution is fully determined by the Prandtl number, the initial, and the boundary conditions.

The distance unit $\sqrt{\varkappa/N}$ is rather short in stellar interiors, and it only weakly depends on the stratification. To see this, we express the Brunt-V\"{a}is\"{a}l\"{a} frequency in terms of the more common stellar-structure parameters,
\begin{equation}
N^2 = \frac{g}{H_\mrm{p}}(\nabla_\mrm{ad} - \nabla),\label{eq:N2}
\end{equation}
where $H_\mrm{p}$ is the pressure scale height, $\nabla_\mrm{ad}$ the adiabatic temperature gradient, and $\nabla$ the actual temperature gradient. Close to a convection zone's boundary, we can write
\begin{equation}
\nabla_\mrm{ad} - \nabla = \alpha\frac{z}{H_\mrm{p}},\label{eq:delta_nabla}
\end{equation}
where $\alpha \approx 10^{-1}$ is a coefficient of proportionality and $z$ the distance from the boundary ($z > 0$ in the stable stratification). When using Eqs.~\ref{eq:N2} and \ref{eq:delta_nabla}, the unit of distance can be expressed as
\begin{equation}
\sqrt{\frac{\varkappa}{N}} = \varkappa^{1/2} \left(\alpha \frac{g}{H_\mrm{p}} \frac{z}{H_\mrm{p}}\right)^{-1/4}, \label{eq:distance_unit}
\end{equation}
which is about $10^7$\,cm for values typical of a point close to the convective/stable interface ($z \approx 10^{-2} H_\mrm{p}$) in the core of a massive ($10\,M_\odot$), main-sequence star ($\varkappa \approx 10^{10}$\,cm$^2$\,s$^{-1}$, $\alpha \approx 10^{-1}$, \mbox{$g \approx 10^5$\,cm\,s$^{-2}$}, $H_\mrm{p} \approx 10^{10}$\,cm).

Two distinct regimes of differential heating can be expected, depending on the amplitude and the spatial scale of the temperature fluctuation imposed on the differentially heated surface. If the heating is strong enough, the heat transport is advection-dominated (i.e. the flow's P\'eclet number is high), and the flow is generally unsteady. A similar phenomenon takes place right at the point where the convective flow leaves the unstable stratification, still retaining some positive temperature fluctuation. It quickly cools down as it rises in the stable medium, its temperature fluctuation turns negative, and the flow is brought to a halt. This is the place where we can impose a lower boundary condition for a much weaker kind of differential-heating-induced flow, in which diffusive heat transport plays a major role (i.e. the flow's P\'eclet number is low). The latter case is the main focus of this paper. We show in Sect.~\ref{sec:results} that such a flow is generally smooth and reaches a stationary state (to be specified in Sect.~\ref{sec:stationarity_structure}) even at rather high values of the Reynolds number, up to $\mrm{Re} = 4 \times 10^3$. This allows us to gain some insight into the problem by exploring the scaling properties of the stationary differential-heating equations, which we do in the next section.

\subsection{Analytical considerations}
\label{sec:analytical_considerations}

The stationary differential-heating problem is described in two dimensions by the set of equations (cf. Eqs.~\ref{eq:continuity2}, \ref{eq:momentum2}, \ref{eq:energy2})
\begin{align}
\pp{u}{x} + \pp{w}{z} &= 0, \label{eq:continuity3} \\
\pp{(uu)}{x} + \pp{(uw)}{z} &= -\pp{p}{x} + \Pr\left(\ppp{u}{x^2} + \ppp{u}{z^2}\right), \label{eq:momentum3x} \\
\pp{(uw)}{x} + \pp{(ww)}{z} &= -\pp{p}{z} + \vartheta + \Pr\left(\ppp{w}{x^2} + \ppp{w}{z^2}\right), \label{eq:momentum3z} \\
\pp{(u\vartheta)}{x} + \pp{(w\vartheta)}{z} &= -w + \ppp{\vartheta}{x^2} + \ppp{\vartheta}{z^2}, \label{eq:energy3}
\end{align}
where $x$ and $z$ are the horizontal and vertical coordinates, respectively, with the $z$ axis pointed against the gravitational acceleration vector, $u$ is the horizontal velocity component, and $w$  the vertical one. In what follows, we show how the characteristic properties of the stationary flow depend on the typical amplitude $\Theta$ and the typical horizontal length scale $L$ of the heating applied.

Assume that there is a well-defined vertical length scale $H$ in the differential heating flow pattern. Let us denote the typical horizontal and vertical velocities by $U$ and $W$, respectively, and the typical pressure fluctuation by $P$. We then introduce a new set of variables $\hat{x}$, $\hat{z}$, $\hat{u}$, $\hat{w}$, $\hat{p}$, and $\hat{\vartheta}$, which all reach values of the order of unity close to the differentially heated surface, and
\begin{align}
x &= L\hat{x}, \label{eq:x_hat} \\
z &= H\hat{z}, \label{eq:z_hat} \\
u &= U\hat{u}, \label{eq:u_hat} \\
w &= W\hat{w}, \label{eq:w_hat} \\
p &= P\hat{p}, \label{eq:p_hat} \\
\vartheta &= \Theta\hat{\vartheta}. \label{eq:theta_hat}
\end{align}
Upon making these substitutions in Eq.~\ref{eq:continuity3}, we obtain
\begin{equation}
\frac{U}{L}\pp{\hat{u}}{\hat{x}} + \frac{W}{H}\pp{\hat{w}}{\hat{z}} = 0,
\end{equation}
which implies the approximate relation
\begin{equation}
\frac{U}{L} \approx \frac{W}{H}. \label{eq:oom_continuity}
\end{equation}
The horizontal momentum equation (Eq.~\ref{eq:momentum3x}) attains the form
\begin{equation}
\pp{(\hat{u}\hat{u})}{\hat{x}} + \pp{(\hat{u}\hat{w})}{\hat{z}} \approx -\frac{P}{U^2}\pp{\hat{p}}{\hat{x}} + \frac{\Pr}{UL}\ppp{\hat{u}}{\hat{x}^2} + \frac{\Pr}{WH}\ppp{\hat{u}}{\hat{z}^2}, \label{eq:oom_momentumx}
\end{equation}
where Eq.~\ref{eq:oom_continuity} has been used, so the equality is only approximate. Nonetheless, we can see that the viscous terms are of the order  of $1/\mrm{Re}_x \equiv \mrm{Pr}/(UL)$ and $1/\mrm{Re}_z \equiv \mrm{Pr}/(WH)$, where $\mrm{Re}_x$ and $\mrm{Re}_z$ are Reynolds-like numbers associated with horizontal and vertical motions, respectively. We introduce this unusual notation to characterise the relative contributions of the two viscous terms in the case of $L \gg H$. We focus on this limit because it turns out to be the relevant one in stellar interiors (see Sect.~\ref{sec:application_to_stars}). From now on, we assume $\mrm{Re}_x \gg 1$ and $\mrm{Re}_z \gg 1$. Equation~\ref{eq:oom_momentumx} shows that pressure fluctuations are of the order of $U^2$ in this high-Reynolds-number limit, so that we can estimate
\begin{equation}
P \approx U^2. \label{eq:oom_p}
\end{equation}
The vertical momentum equation (Eq.~\ref{eq:momentum3z}), with the substitutions defined above and Eqs.~\ref{eq:oom_continuity} and \ref{eq:oom_p}, becomes
\begin{align}
\pp{(\hat{u}\hat{w})}{\hat{x}} + \pp{(\hat{w}\hat{w})}{\hat{z}} \approx \frac{L}{H} \left(-\pp{\hat{p}}{\hat{z}} + \frac{H\Theta}{U^2}\hat{\vartheta}\right) + \frac{1}{\mrm{Re}_x}\ppp{\hat{w}}{\hat{x}^2} + \frac{1}{\mrm{Re}_z}\ppp{\hat{w}}{\hat{z}^2} \label{eq:oom_momentumz}
\end{align}
and implies a close balance between the vertical component of the pressure gradient and the buoyancy-acceleration term provided that $L \gg H$ in addition to $\mrm{Re}_x \gg 1$ and $\mrm{Re}_z \gg 1$. This allows us to estimate
\begin{equation}
U^2 \approx H\Theta, \label{eq:oom_u2}
\end{equation}
which is a plain, order-of-magnitude equality of the characteristic kinetic and potential energies. Finally, the energy equation (Eq.~\ref{eq:energy3}) becomes
\begin{equation}
\pp{(\hat{u}\hat{\vartheta})}{\hat{x}} + \pp{(\hat{w}\hat{\vartheta})}{\hat{z}} \approx -\frac{H}{\Theta}\hat{w} + \frac{1}{UL}\ppp{\hat{\vartheta}}{\hat{x}^2} + \frac{1}{WH}\ppp{\hat{\vartheta}}{\hat{z}^2}. \label{eq:oom_energy}
\end{equation}
The diffusion terms in Eq.~\ref{eq:oom_energy} are of the order of $1/\mrm{Pe}_x \equiv 1/(UL)$ and $1/\mrm{Pe}_z \equiv 1/(WH)$, where $\mrm{Pe}_x$ and $\mrm{Pe}_z$ are P\'eclet-like numbers associated with horizontal and vertical motions, respectively. We introduce them for the very same reason as we did in the case of $\mrm{Re}_x$ and $\mrm{Re}_z$. Making use of Eqs.~\ref{eq:oom_continuity} and \ref{eq:oom_u2}, we can put Eq.~\ref{eq:oom_energy} into the form
\begin{equation}
\pp{(\hat{u}\hat{\vartheta})}{\hat{x}} + \pp{(\hat{w}\hat{\vartheta})}{\hat{z}} \approx \frac{1}{\mrm{Pe}_x} \ppp{\hat{\vartheta}}{\hat{x}^2} + \frac{1}{\mrm{Pe}_z}\left(-\frac{H^{7/2}}{L\Theta^{1/2}}\hat{w} + \ppp{\hat{\vartheta}}{\hat{z}^2}\right), \label{eq:oom_energy2}
\end{equation}
which can be greatly simplified in the double limit of $\mrm{Pe}_x \gg \mrm{Pe}_z$ and $\mrm{Pe}_z \ll 1$. In that case, the two terms in the parentheses on the right-hand side have to closely balance one another, so that we can estimate
\begin{equation}
H \approx \Theta^{1/7} L^{2/7} \label{eq:oom_h}
,\end{equation}
and Eq.~\ref{eq:oom_energy2} becomes linear,
\begin{equation}
\ppp{\hat{\vartheta}}{\hat{z}^2} = \hat{w}. \label{eq:energy4}
\end{equation}
Equation~\ref{eq:energy4} is a special case of the energy equation in the low-P\'eclet-number approximation of \citet{Lignieres99}.

Using Eq.~\ref{eq:oom_h}, we eliminate $H$ from Eq.~\ref{eq:oom_u2} to get an estimate of $U(\Theta, L)$ and, with Eq.~\ref{eq:oom_continuity}, also an estimate of $W(\Theta, L)$. The resulting relations also enable us to express $\mrm{Re}_x$, $\mrm{Re}_z$, $\mrm{Pe}_x$, and $\mrm{Pe}_z$ as functions of $\Theta$, $L$, and $\mrm{Pr}$. This way we obtain
\begin{align}
U &\approx \Theta^{4/7} L^{1/7}, \label{eq:oom_u} \\
W &\approx \Theta^{5/7} L^{-4/7}, \label{eq:oom_w} \\
\mrm{Re}_x &\approx \Theta^{4/7} L^{8/7} \mrm{Pr}^{-1}, \label{eq:oom_rex} \\
\mrm{Re}_z &\approx \Theta^{6/7} L^{-2/7} \mrm{Pr}^{-1}, \label{eq:oom_rez} \\
\mrm{Pe}_x &\approx \Theta^{4/7} L^{8/7}, \label{eq:oom_pex} \\
\mrm{Pe}_z &\approx \Theta^{6/7} L^{-2/7}. \label{eq:oom_pez}
\end{align}

One might be tempted to estimate the time scale $\tau$ of flow acceleration towards the stationary state directly from the buoyancy acceleration $\Theta$ provided by the temperature fluctuation imposed on the bottom boundary. It is crucial to realise that, as Eq.~\ref{eq:oom_momentumz} shows, the buoyancy acceleration is almost completely compensated for by the vertical component of the pressure gradient in the case $L \gg H$. It is only their difference that contributes to the vertical acceleration. We can, however, consider the horizontal acceleration provided by the horizontal component of the pressure gradient and write $U/\tau \approx P/L \approx U^2/L$ (see Eq.~\ref{eq:oom_p}). Using Eq.~\ref{eq:oom_u} we obtain
\begin{equation}
\tau \approx \Theta^{-4/7} L^{6/7}. \label{eq:oom_tau}
\end{equation}
Finally, we would like to point out that the characteristic thermal-diffusion length scale corresponding to the time scale $\tau$ is $\tau^{1/2} \approx \Theta^{-2/7} L^{3/7}$, which scales with $\Theta$ and $L$ in quite a different way than $H$ does (see Eq.~\ref{eq:oom_h}). This comes about because our estimates take the back reaction of the flow on the temperature distribution into account.

\subsection{Numerical solutions}
\label{sec:numerical_solutions}

\begin{table}
\begin{center}
\begin{tabular}{l c c r@{$\,\times\,$}l r@{$\,\times\,$}l}
\hline\hline
Id. & $\Theta$ & $L$ & \multicolumn{2}{c}{$\mrm{Pe}_x$} & \multicolumn{2}{c}{$\mrm{Pe}_z$} \rule[-0.9ex]{0pt}{3.25ex} \\
\hline
sr00 & $10^{0}$ & $10^{1}$ & 8.5&$10^{0}$ & 2.5&$10^{0}$ \rule{0pt}{2.25ex} \\
sr01 & $10^{0}$ & $10^{2}$ & 1.4&$10^{2}$ & 1.3&$10^{0}$ \\
sr02 & $10^{0}$ & $10^{3}$ & 2.0&$10^{3}$ & 5.5&$10^{-1}$ \\
sr03 & $10^{0}$ & $10^{4}$ & 2.9&$10^{4}$ & 2.7&$10^{-1}$ \\
sr10 & $10^{-1}$ & $10^{1}$ & 2.5&$10^{0}$ & 3.1&$10^{-1}$ \\
sr11 & $10^{-1}$ & $10^{2}$ & 4.0&$10^{1}$ & 1.4&$10^{-1}$ \\
sr12 & $10^{-1}$ & $10^{3}$ & 5.5&$10^{2}$ & 6.8&$10^{-2}$ \\
sr13 & $10^{-1}$ & $10^{4}$ & 7.7&$10^{3}$ & 3.5&$10^{-2}$ \\
sr20 & $10^{-2}$ & $10^{1}$ & 7.2&$10^{-1}$ & 4.1&$10^{-2}$ \\
sr21 & $10^{-2}$ & $10^{2}$ & 1.1&$10^{1}$ & 1.8&$10^{-2}$ \\
sr22 & $10^{-2}$ & $10^{3}$ & 1.5&$10^{2}$ & 9.3&$10^{-3}$ \\
sr23 & $10^{-2}$ & $10^{4}$ & 2.1&$10^{3}$ & 4.8&$10^{-3}$ \\
sr30 & $10^{-3}$ & $10^{1}$ & 2.0&$10^{-1}$ & 5.4&$10^{-3}$ \\
sr31 & $10^{-3}$ & $10^{2}$ & 2.9&$10^{0}$ & 2.5&$10^{-3}$ \\
sr32 & $10^{-3}$ & $10^{3}$ & 4.0&$10^{1}$ & 1.3&$10^{-3}$ \\
sr33 & $10^{-3}$ & $10^{4}$ & 5.5&$10^{2}$ & 6.7&$10^{-4}$ \\
\hline
\end{tabular}
\caption{Parameters of the series of simulations sampling a patch of the parameter space $\{\Theta,\,L\}$ at the constant value of $\mrm{Re} = 2.6 \times 10^2$.}
\label{tab:sr_models}
\end{center}
\end{table}
\begin{table}
\begin{center}
\begin{tabular}{l r@{$\,\times\,$}l r@{$\,\times\,$}l}
\hline\hline
Id. & \multicolumn{2}{c}{Resolution} & \multicolumn{2}{c}{$\mrm{Re}$} \rule[-0.9ex]{0pt}{3.25ex} \\
\hline
Re32 & 32&32 & 3.2&$10^{1}$ \rule{0pt}{2.25ex} \\
Re64 & 64&64 & 6.4&$10^{1}$ \\
Re128 & 128&128 & 1.3&$10^{2}$ \\
Re256 & 256&256 & 2.6&$10^{2}$ \\
Re512 & 512&512 & 5.1&$10^{2}$ \\
Re1024 & 1024&1024 & 1.0&$10^{3}$ \\
Re2048 & 2048&2048 & 2.0&$10^{3}$ \\
Re4096 & 4096&4096 & 4.1&$10^{3}$ \\
\hline
\end{tabular}
\caption{Parameters of the series of simulations with Re increasing at the fixed values of $\Theta = 10^{-3}$ and $L = 10^0$.}
\label{tab:re_models}
\end{center}
\end{table}

The order-of-magnitude estimates derived in the preceding section assume that the flow is stationary and that there is a well-defined vertical length scale in the flow pattern. We performed a series of time-dependent, numerical simulations of the differential heating problem to confirm these assumptions and to determine how the solutions depend the Reynolds number and how they decrease with height.

We have developed a specialised code dedicated to the study of the differential-heating problem, because the problem places rather high demands on the numerical scheme. For instance, it has to tackle the highly diffusive nature of the flow and its high aspect ratio and resolve a wide dynamic range within a single simulation box. The code is of the finite-difference type, and it solves the differential-heating equations on a collocated grid using a variant of the MacCormack integration scheme. The Poisson equation for pressure, which can be derived from Eqs.~\ref{eq:continuity2} and \ref{eq:momentum2} (or \ref{eq:momentum4}, see below), is solved by a spectral method. Heat-diffusion terms are treated implicitly, again by a spectral method. In what follows, we discuss a few selected issues related to the numerical solution of the differential-heating equations that need to be borne in mind when interpreting our results. The reader interested in the details of the numerical scheme is referred to App.~\ref{sec:numerical_methods}.

We use periodic boundaries in the horizontal direction and force the shear stress and the vertical velocity component to vanish at the lower and upper boundaries of the computational domain. One could also use non-slip boundaries, but these are hardly more akin to the physical reality that motivated this study in the first place, so we omit this case. We impose a temperature fluctuation in the form $\vartheta(x, 0) = \Theta\sin(\pi x/L)$ at the bottom boundary and force the temperature fluctuation to vanish at the upper boundary. The parameters $\Theta$ and $L$ can be identified with the same symbols as introduced in Sect.~\ref{sec:analytical_considerations}.

The high thermal diffusivity in the differential-heating problem forces us to use long implicit time steps for the heat-diffusion terms, which might have an adverse effect on the accuracy of the results. To show that this is not the case, we re-computed the simulations sr03, sr30, and Re1024 (see Tables~\ref{tab:sr_models} and \ref{tab:re_models} and Sect.~\ref{sec:results}), decreasing the time step by a factor of ten. This brings about a change in the velocity field, which is of the order of 0.1\% in the cases sr03 and sr30 and of the order of 1\% in the case of Re1024 (measured well away from the field's zeroes). The reason for this insensitivity to the time step is the low P\'eclet number of the flow. \cite{Lignieres99} shows that in the low-Pe regime, the energy equation can be approximated by a Poisson equation for the temperature fluctuation with $w$ as a source term (see also our Eq.\ref{eq:energy4}). We do not use this approximation to make our code more versatile; instead, we naturally obtain a close equilibrium between the terms $\bm \nabla^2 \vartheta$ and $w$ in Eq.~\ref{eq:energy2} when the P\'eclet number is low. This equilibrium is reached so quickly that details of the evolution of $\vartheta$ towards the equilibrium become irrelevant.

It is a well-known fact that any numerical advection scheme either requires adding a so-called artificial-viscosity term to guarantee stability or it involves some viscous behaviour implicitly. In either case, the effective Reynold number does not even come close to the astrophysical regime with current computing facilities. The artificial viscosity (be it explicit or implicit) thus exceeds the physical one by a wide margin, so it demands special attention.

Suppose we include an explicit viscous term as in Eq.~\ref{eq:momentum2} to model the artificial viscosity. Equations~\ref{eq:oom_rex} and \ref{eq:oom_rez} show that for $L \approx 10^3$ (equivalent to $\sim H_p$ in the astrophysical case mentioned in Sect.~\ref{sec:problem_formulation}) we have $\mrm{Re}_z \approx 10^{-4}\,\mrm{Re}_x$ as a consequence of $H \ll L$. Using equidistant grids with up to $10^3$ grid points in each direction, we can achieve \mbox{$\mrm{Re}_x \approx 10^3$}. It follows that \mbox{$\mrm{Re}_z \lesssim 10^{-1}$} and the vertical momentum transport is dominated by the artificial-viscosity term. A value $\mrm{Re}_z \gg 1$ is, however, expected in stellar interiors. We use a simple workaround, replacing the viscous term $\mrm{Pr}\nabla^2 \bm u$ by the anisotropic form $\mrm{Pr}_x\,\partial^2\bm u / \partial x^2 + \mrm{Pr}_z\,\partial^2\bm u / \partial z^2$. The coefficients $\mrm{Pr}_x$ and $\mrm{Pr}_z$ are re-computed at each time step from the relations \mbox{$\mrm{Pr}_x = h_x \max |u|\, \mrm{Re}_\mrm{grid}^{-1}$} and $\mrm{Pr}_z = h_z \max |w|\, \mrm{Re}_\mrm{grid}^{-1}$, where $h_x$ and $h_z$ are the horizontal and vertical grid spacings, respectively, and $\mrm{Re}_\mrm{grid}$ is the Reynolds number on the grid scale. We performed a few numerical tests of the code on a convection problem to determine that the value $\mrm{Re}_\mrm{grid} = 4$ is a conservative trade-off between the amount of viscous dissipation and the code's stability, so we use this value in all the simulations presented here.

The anisotropic form of artificial viscosity enables us to reach $\mrm{Re}_x \gg 1$ and $\mrm{Re}_z \gg 1$ at the same time on a grid of reasonable size. We show in Sect.~\ref{sec:results} that the solutions with constant values of $\mrm{Pr}_x$ and $\mrm{Pr}_z$ decay exponentially with height. The effective, local Reynolds number decreases in proportion to the flow speed, and the solutions quickly become dominated by the artificial viscosity. This would also happen in an (otherwise idealised) stellar interior at some point but that point would be much farther from the convection zone's boundary. Therefore, we generalise the artificial-viscosity terms, and the momentum equation (Eq.~\ref{eq:momentum2}) in 2D becomes
\begin{equation}
\DD{\bm u}{t} = - \bm\nabla p + \vartheta\bm k + \pp{}{x}\left[\mrm{Pr}_x(z) \pp{\bm u}{x}\right] + \pp{}{z}\left[\mrm{Pr}_z(z) \pp{\bm u}{z}\right], \label{eq:momentum4}
\end{equation}
where are $\mrm{Pr}_x(z)$ and $\mrm{Pr}_z(z)$ are proportional to $\mrm{e}^{-\eta z}$ with $\eta$ being an adjustable parameter. We set $\eta = 0$ when we are not interested in the precise vertical profiles  and use $\eta > 0$ to suppress the viscous terms when examining how the solutions decrease with height. The latter case, $\eta > 0$, is a rather touchy problem because one has to increase $\eta$ in a few steps, always using the (almost) stationary flow from a previous run as an initial condition for the next run. Overestimating the value of $\eta$ can lead to a lack of viscous dissipation in some parts of the computational domain and a numerical instability ensues. Finally, the very goal that we want to achieve by this treatment, i.e. the flow dynamics' being dominated by inertial terms at great heights, becomes an issue since such a flow evolves on the extremely long time scale corresponding to its low speed.

\section{Results}
\label{sec:results}

\subsection{The stationarity and structure of the flow}
\label{sec:stationarity_structure}

Our numerical investigation of the diffusion-dominated differential heating problem reveals that the flow reaches a stationary state at all values of the Reynold number that we have been able to achieve (up to $\mrm{Re} = 4\times 10^3$). We use the rate of change of the quantity $u_\mrm{max} = \max |u|$ (taken over the whole simulation box) as a convergence monitor and a basis of our criterion for deciding the flow's stationarity. We show in Sect.~\ref{sec:analytical_considerations} that the relevant dynamical time scale near the differentially heated boundary should be close to $\tau$ given by Eq.~\ref{eq:oom_tau} (confirmed a posteriori, see below). We pronounce the flow stationary and stop the simulation once the condition
\begin{equation}
\left|\frac{1}{u_{max}}\frac{\partial u_\mrm{max}}{\partial t}\,\tau\right| < 10^{-3}
\label{eq:stationarity_condition}
\end{equation}
has been met at least for one time scale $\tau$. A direct implementation of this condition would involve extrapolation from the time scale of one time step, $\Delta t$, to a much longer time scale $\tau$, which would amplify the round-off noise by a factor of $\tau/\Delta t \gg 1$. Instead, we approximate Eq.~\ref{eq:stationarity_condition} by
\begin{equation}
\left|\frac{1}{u_{max}}\frac{u_\mrm{max} - \overline{u_\mrm{max}}}{\tau}\,\tau\right| < 10^{-3},
\label{eq:stationarity_condition2}
\end{equation}
where $\overline{u_\mrm{max}}$ is the Euler-time-stepped solution of the equation \mbox{$\partial \overline{u_\mrm{max}} / \partial t = (u_\mrm{max} - \overline{u_\mrm{max}})/\tau$}. Thus, $\overline{u_\mrm{max}}$ is a smoothed version of $u_\mrm{max}$, lagging behind it approximately by $\tau$ in time.

The flow in all of the simulations presented in this paper is composed of several layers of overturning cells with flow speed rapidly decreasing from one layer to the next (see Figs.~\ref{fig:flow_structure} or \ref{fig:decreasing_viscosity}). We characterise the flow properties close to the differentially heated surface by a vertical length scale $H$, defined as the height above the hottest spot at which the flow first turns over (i.e. $w(L/2, H) = 0$), and the typical horizontal and vertical velocity components $U = \frac{1}{2}\max(u)$ and $W = \frac{1}{2}\max(w)$, respectively, where the maxima are taken over the whole simulation box. The symbols $H$, $U$, and $W$ can be identified with the same symbols as used in Sect.~\ref{sec:analytical_considerations}. The flow always reaches its maximal horizontal speed at the bottom boundary and the maximal vertical speed in the first overturning cell above the hot spot. The flow pattern is asymmetric, with the maximum downward flow speed (reached above the cold spot), $\max(-w)$, always lower than the maximal upward flow speed, $\max(w)$. We define the characteristic numbers $\mrm{Re}_x$, $\mrm{Re}_z$, $\mrm{Pe}_x$, and $\mrm{Pe}_z$ in an analogous way to what is used in Sect.~\ref{sec:analytical_considerations} with the difference that now we have two Prandtl-like numbers $\mrm{Pr}_x$ and $\mrm{Pr}_z$ instead of one Prandtl number $\mrm{Pr}$.

\subsection{Scaling relations}

\begin{figure*}
\centering
\includegraphics[width=16cm, height=23cm]{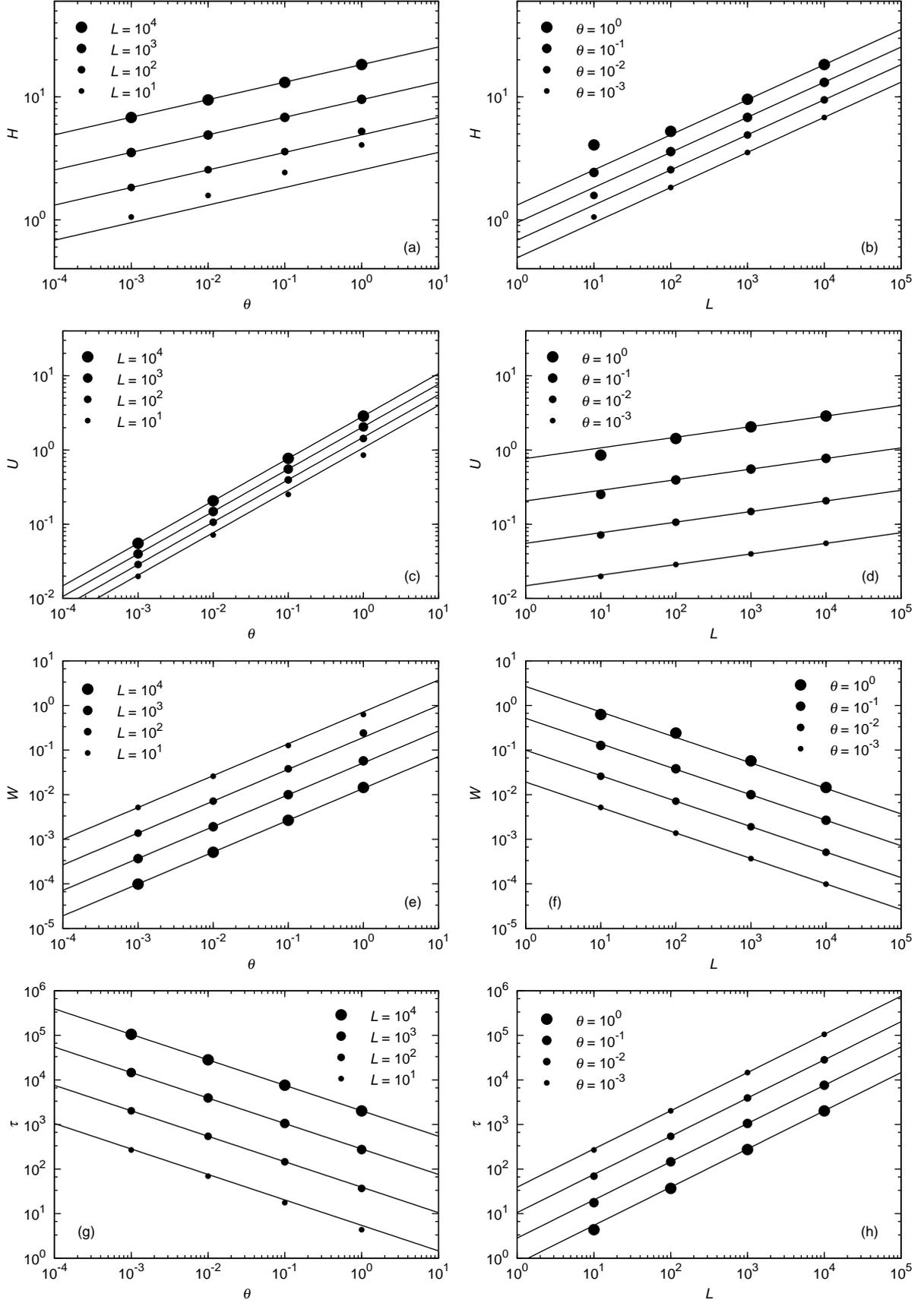}
\caption{Dependence of the global flow characteristics on the heating amplitude $\Theta$ and length scale $L$. Circles show the values derived from numerical simulations (Table~\ref{tab:sr_models}). Solid lines show the scaling relations (Eqs.~\ref{eq:oom_h}, \ref{eq:oom_u}, \ref{eq:oom_w}, and \ref{eq:oom_tau}), normalised to fit all but the four simulations at $L=10^1$, which are expected to deviate from the scaling relations.}
\label{fig:scaling_relations}
\end{figure*}

\begin{figure*}
\centering
\includegraphics[height=8.6cm]{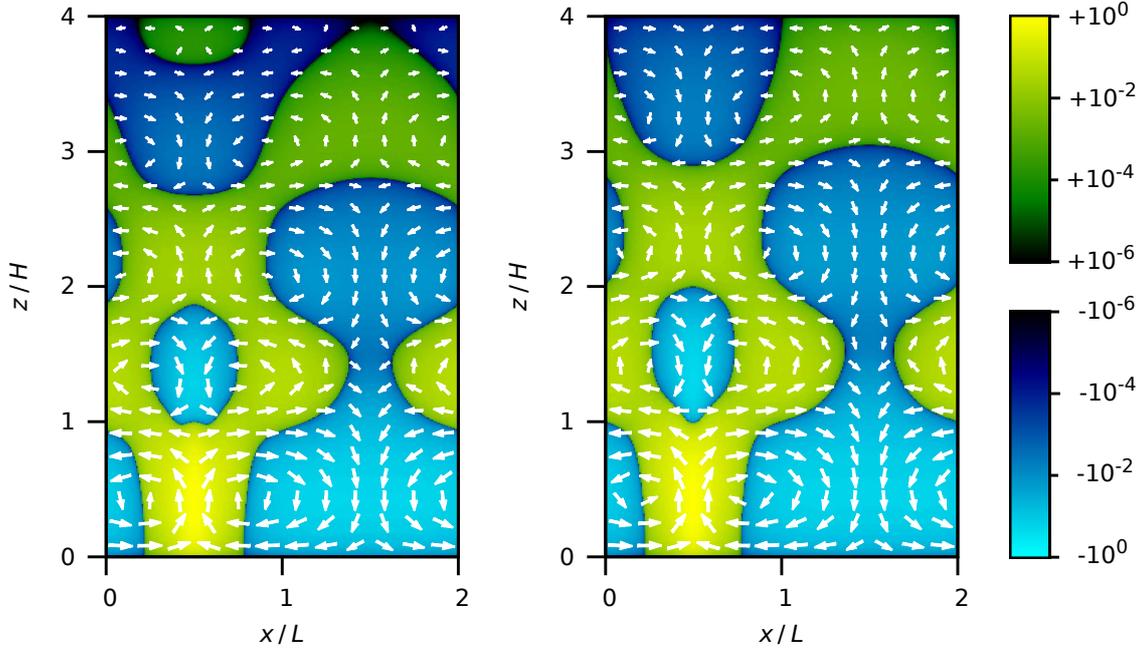}
\caption{Comparison of the flow structure in two simulations with very disparate heating parameters. The left panel shows simulation sr30 (\mbox{$\Theta = 10^{-3}$}, $L = 10^1$). The right panel shows simulation sr03 ($\Theta = 10^0$, $L = 10^4$). In both cases, the vertical velocity component $w$, normalised to its maximal absolute value, is plotted on a split logarithmic colour scale. The length of the velocity vectors (arrows) is scaled in a non-linear way to aid visualisation. The spatial coordinates are normalised by the characteristic length scales defined in Sects.~\ref{sec:numerical_solutions} and \ref{sec:results}.}
\label{fig:flow_structure}
\end{figure*}

We computed a grid of 16 simulations to verify the analytical relations derived in Sect.~\ref{sec:analytical_considerations}. All of these simulations, summarised in Table~\ref{tab:sr_models}, have a resolution of \mbox{$256 \times 512$}, and the vertical grid spacing was adjusted so as to obtain \mbox{$\mrm{Re}_x = \mrm{Re}_z \equiv \mrm{Re} = 2.6 \times 10^2$}. The decision to fix the value of $\mrm{Re}$ is motivated by the fact that the flow pattern turns out to be scalable over a large part of the parameter space provided that $\mrm{Re} = \mrm{const.}$. In other words, while changing the heating parameters $\Theta$ and $L$ at $\mrm{Re} = \mrm{const.}$ does change the amplitude and the vertical scale of the flow, the structure of the flow, as seen in a system of normalised coordinates $x/L$ and $z/H$, remains unchanged (see Fig.~\ref{fig:flow_structure}). We show in Fig.~\ref{fig:scaling_relations} our numerical results as compared with the scaling relations fitted to all but the four data points at $L = 10^1$. The excluded data points do not comply well with the premise $L \gg H$ and are thus expected not to follow the scaling relations. Allowing only the constants of proportionality to change in the fitting process, we obtain
\begin{align}
H &= 1.3\, \Theta^{1/7} L^{2/7}, \label{eq:h_fit} \\
U &= 0.77\, \Theta^{4/7} L^{1/7}, \label{eq:u_fit} \\
W &= 2.7\, \Theta^{5/7} L^{-4/7}, \label{eq:w_fit} \\
\tau &= 0.76\, \Theta^{-4/7} L^{6/7}. \label{eq:tau_fit}
\end{align}
The unexpectedly good fit is a result of the flow's scalability.
\begin{figure}
\centering
\includegraphics[width=88mm]{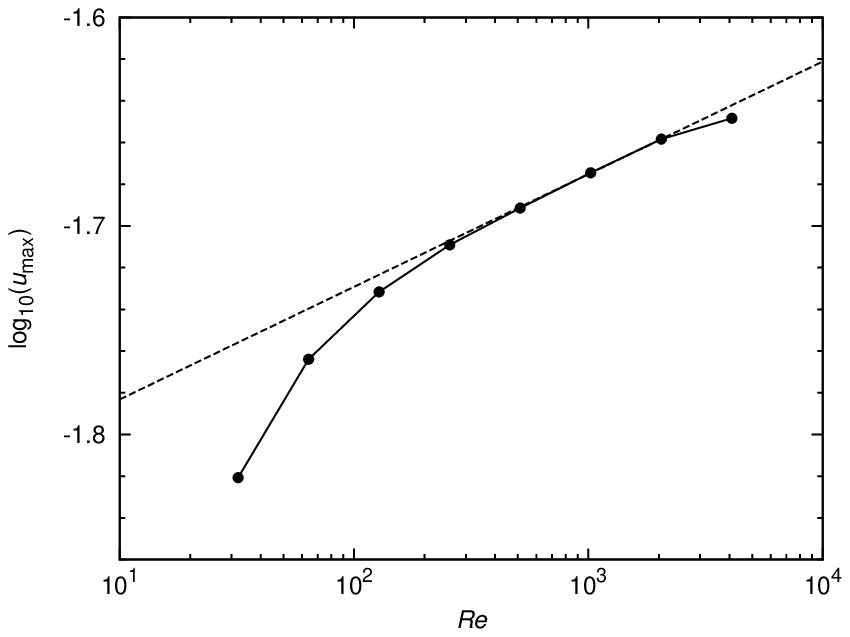}
\caption{Dependence of the maximal horizontal velocity on the Reynolds number. Simulation data (circles) are connected by the solid line to guide the eye. The scaling law $u_\mrm{max} \propto Re^{0.054}$ is shown by the dashed line for comparison.}
\label{fig:u_max_Re}
\end{figure}
\begin{figure}
\centering
\includegraphics[width=88mm]{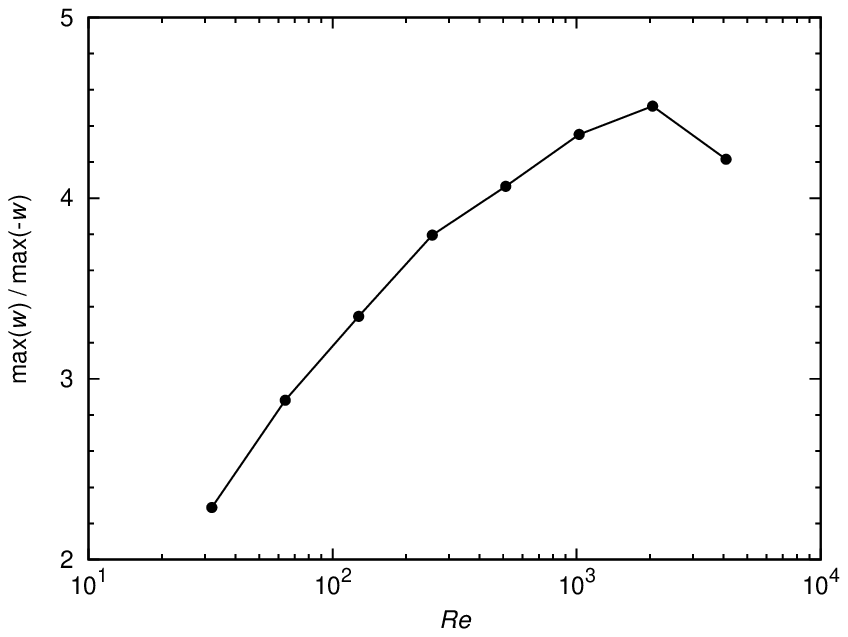}
\caption{Dependence on the Reynold number of the flow's asymmetry, characterised by the ratio of the maximum upward and downward flow speeds.}
\label{fig:flow_asymmetry}
\end{figure}

The constants of proportionality in Eqs.~\ref{eq:h_fit}\,--\,\ref{eq:tau_fit}, as well as the structure of the flow, depend on $\mrm{Re}$. We illustrate this dependence by computing a series of simulations with resolution ranging from $32^2$ to $4096^2$. This way, we cover about two orders of magnitude in the Reynolds number from $3 \times 10^1$ to $4 \times 10^3$ (again with $\mrm{Re}_x = \mrm{Re}_z \equiv \mrm{Re}$). We perform this experiment at $L = \mrm{const.}$ and $\Theta = \mrm{const.}$, so any change in $\mrm{Re}$ reflects a change in $\mrm{Pr}_x$ and $\mrm{Pr}_z$. Nevertheless, we present the dependence on $\mrm{Re}$, because the scalability of the flow shows that the absolute values of $\mrm{Pr}_x$ and $\mrm{Pr}_z$ do not matter. Ideally, we should choose the heating parameters so as to have $\Theta \ll 1$ and $L \gg 1$ as the scaling relations hold true in this limit (see Sect.~\ref{sec:analytical_considerations}). 

Equation~\ref{eq:tau_fit} shows, however, that the flow's dynamical time scale becomes extremely long  in the same limit, thus making any high-resolution computation unfeasible. Therefore we use \mbox{$\Theta = 10^{-3}$} and $L = 10^{0}$, which still keeps the energy equation approximately linear (since $\mrm{Pe}_x \approx \mrm{Pe}_x \approx 10^{-2}$), but we forgo having $L \gg H$. Nevertheless, we expect the changes in the flow with increasing $\mrm{Re}$ in this case to be similar to those that would be seen in a simulation with $L \gg H$ because of the energy equation's being linear in both cases. All of this series of simulations, summarised in Table~\ref{tab:re_models}, reach the stationary state as defined by Eq.~\ref{eq:stationarity_condition2}. Figure~\ref{fig:u_max_Re} shows that the maximum horizontal velocity in the computational domain slowly increases in proportion to $\mrm{Re}^{0.054}$ in the high-Re regime. The flow also becomes increasingly asymmetric, as shown by the ratio of the maximum upward and downward flow speeds plotted as a function of Re in Fig.~\ref{fig:flow_asymmetry}. The seemingly asymptotic trend changes at the highest Reynolds number considered, but we do not know the reason for this change.

\subsection{Flow at great heights}
\label{sec:flow_at_great_heights}

\begin{figure}
\centering
\includegraphics[width=88mm]{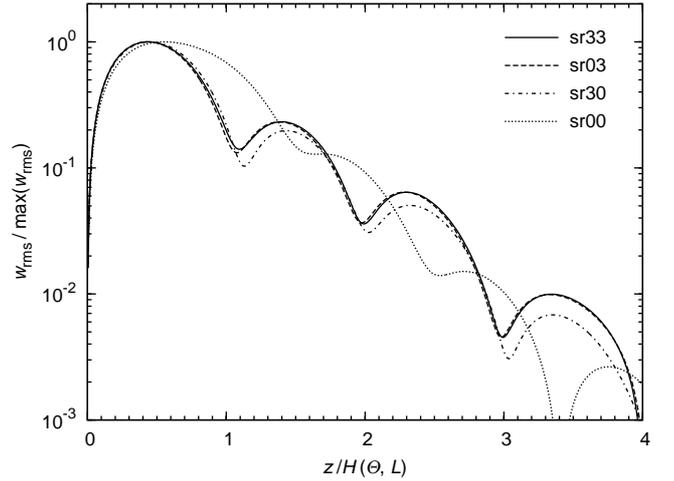}
\caption{Decline with height of the relative rms vertical velocity in four simulations with very different heating parameters. The flows are reasonably close to a stationary state only up to $z/H \doteq 2.5$.}
\label{fig:rms_w1}
\end{figure}

The flow speed in all our simulations quickly decreases with height. Figure~\ref{fig:rms_w1} compares the vertical profiles of the root-mean-square (rms; computed in the $x$ direction) vertical velocity component, $w_\mrm{rms}(z)$, in four simulations with widely disparate heating parameters (sr00, sr03, sr30, and sr33). We find that $w_\mrm{rms}$ decreases approximately as $e^{-\beta_w z/H}$ in a global sense with $\beta_w \doteq 1.5$ almost independently of $\Theta$ and $L$. We use the values $H(\Theta,\,L)$ given by Eq.~\ref{eq:h_fit} instead of those measured in the simulations to normalise the $z$ coordinate, because this brings the slopes much closer to one another. We have to keep in mind, though, that these flows are reasonably close to a stationary state only up to $z/H \doteq 2.5$, because our convergence criterion (Eq.~\ref{eq:stationarity_condition2}) is ignorant of the weak flow in the upper part of the simulation box, and consequently, that part of the flow is still slowly evolving when the computation is stopped.

\begin{figure*}
\centering
\includegraphics[height=8.6cm]{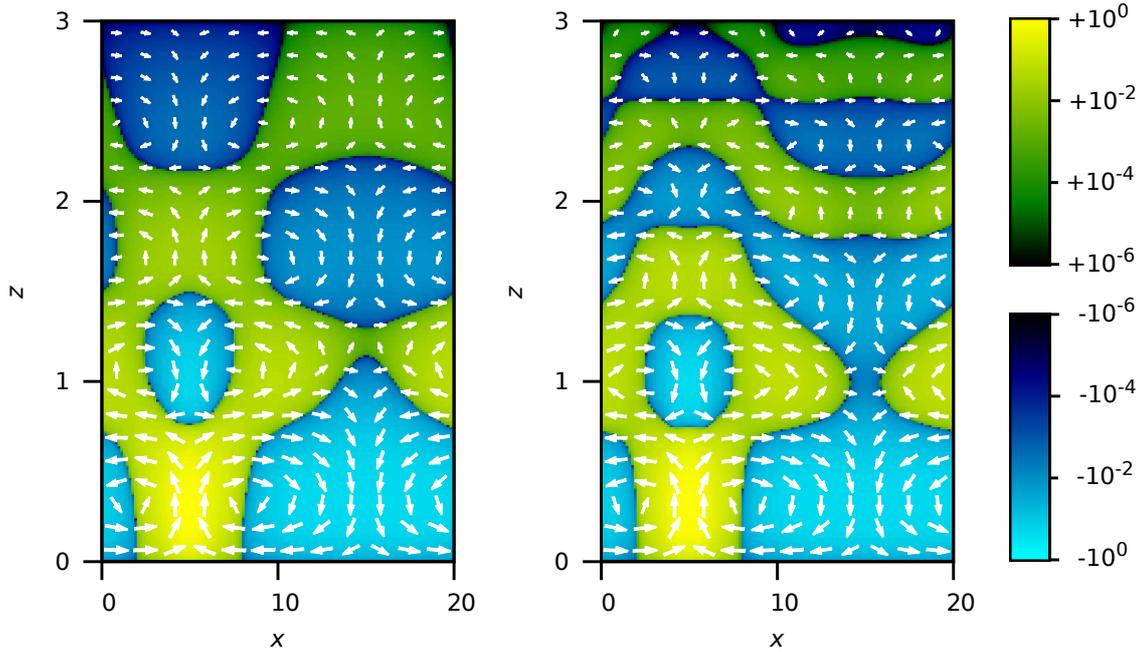}
\caption{Effect of two different artificial-viscosity prescriptions on the flow structure. The constant-Pr flow ($Pr_x, Pr_z = \mrm{const.}$; left panel) is compared with the constant-Re flow ($Pr_x, Pr_z \propto \mrm{e}^{-2z}$; right panel). The vertical velocity component $w$ is in both cases plotted on a split-logarithmic colour scale. The length of the velocity vectors (arrows) is scaled in a non-linear way to aid visualisation.}
\label{fig:decreasing_viscosity}
\end{figure*}

\begin{figure}
\centering
\includegraphics[width=88mm]{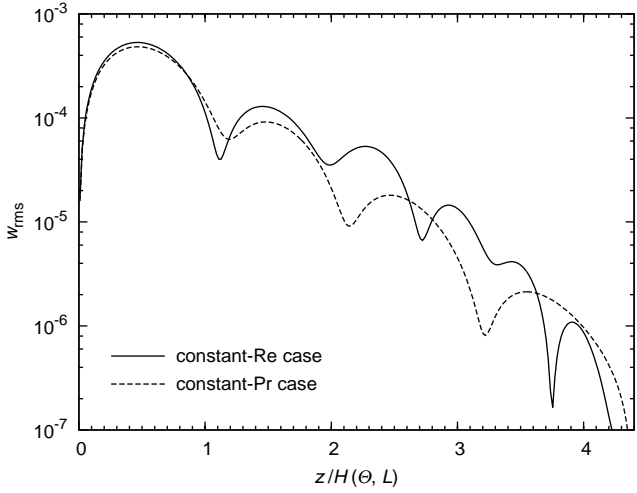}
\caption{Effect of two different artificial-viscosity prescriptions on the rms vertical velocities. Note that the constant-Pr flow can be considered stationary over the whole range shown whereas the constant-Re flow is stationary only up to $z/H \doteq 3.8$.}
\label{fig:rms_w2}
\end{figure}

\begin{figure}
\centering
\includegraphics[width=88mm]{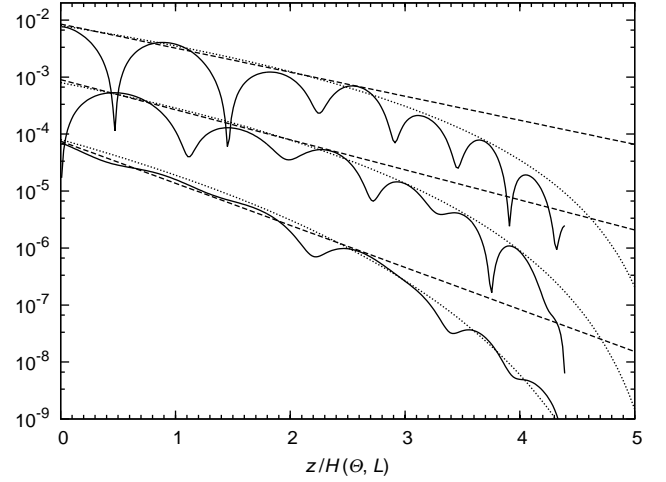}
\caption{Comparison of the rms velocities and temperature fluctuations in the constant-$\mrm{Re}$ case with two models approximating their global behaviour. Solid lines show $u_\mrm{rms}$ (top), $w_\mrm{rms}$ (middle), and $\vartheta_\mrm{rms}$ (bottom). Dashed lines show the model, in which $u_\mrm{e}(z) \propto e^{-\beta_u z/H}$, $w_\mrm{e}(z) \propto e^{-\beta_w z/H}$, and $\vartheta_\mrm{e}(z) \propto e^{-\beta_\vartheta z/H}$ with $\beta_u = \frac{4}{7}\beta_\vartheta$, $\beta_w = \frac{5}{7}\beta_\vartheta$, and $\beta_\vartheta = 1.7$. Dotted lines show the improved model given by Eqs.~\ref{eq:theta_e2}, \ref{eq:u_e}, and \ref{eq:w_e} with \mbox{$\gamma = 1.3$}. The coefficients of proportionality have been adjusted for each variable independently.}
\label{fig:rms_tuw}
\end{figure}

The simulations discussed so far use constant artificial-viscosity parameters $\mrm{Pr}_x$ and $\mrm{Pr}_z$, which leads to a rapid decrease in the local Reynolds number with height (in proportion to the decreasing flow speed). We computed another two simulations, this time with $\Theta=10^{-4}$ and $L=10^1$. In the first one, we set $\mrm{Pr}_x = \mrm{const.}$ and $\mrm{Pr}_z = \mrm{const.}$ (the constant-Pr case hereinafter), just as we have done so far. In the other one, we set $\mrm{Pr}_x \propto e^{-\eta z}$ and $\mrm{Pr}_z \propto e^{-\eta z}$ as described in Sect.~\ref{sec:numerical_solutions} to keep a local version of the Reynolds number approximately constant (the constant-Re case hereinafter). We increased the slope $\eta$ from 0 in a few steps in order to make the ratio of the rms advection terms to the rms viscous terms, i.e. the local Reynolds number, as independent of height as possible; $\eta = 2$ turns out to be a good compromise in this case. There is a large-scale, residual variation by about a factor of four in the local Reynolds number, because the simple exponential profile of the artificial viscosity is not flexible enough to compensate for it. Using Fig.~\ref{fig:u_max_Re} we estimate that this variation can change the velocities by $\sim 0.1$\,dex at most. Since our usual stopping condition, Eq.~\ref{eq:stationarity_condition2}, cannot ``sense'' the weak flow at great heights, we judge the stationarity of the flow by comparing the rms values of the $\partial/\partial t$ terms to the rms values of all other terms that appear in Eq.~\ref{eq:momentum4} and require the former to be significantly smaller than the latter. This way, we obtain the results summarised in Figs.~\ref{fig:decreasing_viscosity}, \ref{fig:rms_w2}, and \ref{fig:rms_tuw}. The constant-Pr flow can be considered stationary over the whole range shown, whereas the constant-Re flow is
only stationary for $z/H \lesssim 3.8$, because the topmost part of that flow evolves so slowly that a global oscillation develops before it has reached equilibrium (see Sect.~\ref{sec:late-time_evolution} for details). 

Figure~\ref{fig:rms_w2} illustrates that the flow is somewhat faster at $z/H > 1$ in the constant-Re case, as could be expected from the massive increase in the local Reynolds number by as much as two orders of magnitude at $z \doteq 3$. Much more interesting is, however, that the overturning cells in the constant-Re case become apparently thinner with increasing height, hence with decreasing local temperature fluctuation. This observation suggests that the scaling relations derived in Sect.~\ref{sec:analytical_considerations} could be used locally (see the dependence of $H$ on $\Theta$ in Eq.~\ref{eq:oom_h}). Another piece of evidence for this hypothesis is shown in Fig.~\ref{fig:rms_tuw}, in which we compare the relative rates of decrease in $\vartheta_\mrm{rms}(z)$, $u_\mrm{rms}(z),$ and $w_\mrm{rms}(z)$. The envelope of $\vartheta_\mrm{rms}(z)$ can be approximated well by the function $\vartheta_\mrm{e}(z) \propto e^{-\beta_\vartheta z/H}$ with $\beta_\vartheta = 1.7$ for $z/H \lesssim 3$. We then regard $\vartheta_\mrm{e}(z)$ as an estimate of the local temperature fluctuation and rewrite the scaling relations for the velocity components, Eqs.~\ref{eq:oom_u} and \ref{eq:oom_w}, to obtain their local versions,
\begin{align}
u_\mrm{e}(z) &\approx \vartheta_\mrm{e}(z)^{4/7} L^{1/7}, \label{eq:oom_u_local} \\
w_\mrm{e}(z) &\approx \vartheta_\mrm{e}(z)^{5/7} L^{-4/7}, \label{eq:oom_w_local}
\end{align}
where $u_\mrm{e}(z)$ and $w_\mrm{e}(z)$ are expected to be good envelope models of $u_\mrm{rms}(z)$ and $w_\mrm{rms}(z)$. In other words, we expect $u_\mrm{e}(z) \propto e^{-\beta_u z/H}$ and $w_\mrm{e}(z) \propto e^{-\beta_w z/H}$ with $\beta_u = \frac{4}{7}\beta_\vartheta$ and $\beta_w = \frac{5}{7}\beta_\vartheta$. Indeed, these scalings turn out to be correct, as shown in Fig.~\ref{fig:rms_tuw}. Similarly, we can produce a local version of Eq.~\ref{eq:oom_h},
\begin{equation}
h(z) \approx \vartheta_\mrm{e}(z)^{1/7} L^{2/7}, \label{eq:oom_h_local}
\end{equation}
where $h(z)$ is a local, height-dependent estimate of a vertical length scale analogous to $H$. As a result, we expect $h(z) \propto e^{-\beta_h z/H}$ with $\beta_h = \frac{1}{7}\beta_\vartheta$, i.e. a slow thinning of the overturning cells with increasing height, similar to what we observe in Figs.~\ref{fig:decreasing_viscosity}, \ref{fig:rms_w2}, and \ref{fig:rms_tuw}. This seemingly innocuous phenomenon has very grave consequences for the flow at great heights. Instead of fading out exponentially, it decreases even faster (see Fig.~\ref{fig:rms_tuw}). We expand on this in Sect.~\ref{sec:improving_the_model} and derive a better model for the flow's decline with height to show that the flow speed drops dramatically above a certain point.

\subsection{Late-time evolution of the flow}
\label{sec:late-time_evolution}

Having continued some of our simulations for as much as $10^4 \tau$, we discover an intriguing phenomenon. At first, a horizontal mean shear flow develops on top of the differential-heating flow. Its amplitude grows, and the shear flow begins to oscillate at some point. Finally, the oscillation saturates at an amplitude ranging from $\sim 10^{-3}$ to $\sim 10^0$ of the differential heating flow's amplitude, depending on the parameters of the simulation. The oscillation's period and development time strongly decrease with increasing Reynolds number. They do not seem to have an upper limit but approach $10\tau$ at $\mrm{Re} \approx 10^3$. This phenomenon most likely has a physical origin because decreasing the time step by a factor of ten does not affect the shear flow or its behaviour significantly. Any detailed study of this phenomenon is certainly beyond the scope of this paper, but our preliminary research suggests that it is unlikely to be a cumulative effect induced by internal gravity waves since it (1) also occurs in very small computational boxes, in which all internal-wave modes are over-damped by radiative diffusion, and (2) the temporal spectra of the average horizontal velocity are featureless at periods significantly shorter than that of the shear flow oscillation.

\section{Interpretation of the results}

\subsection{Improving the model at great heights}
\label{sec:improving_the_model}

After picking up the threads of Sect.~\ref{sec:flow_at_great_heights}, we presently
find that the diffusion-dominated, high-Re differential-heating flow actually decreases faster than exponentially with height. To see this, we make use of two results from Sect.~\ref{sec:flow_at_great_heights}. First, that the scaling relations derived in Sect.~\ref{sec:analytical_considerations} have their local analogues, which hold within the flow (compare Eqs.~\ref{eq:oom_h}, \ref{eq:oom_u}, and \ref{eq:oom_w} with Eqs.~\ref{eq:oom_h_local}, \ref{eq:oom_u_local}, and \ref{eq:oom_w_local}, respectively). Second, that the envelope of $\vartheta_\mrm{rms}(z)$ can be approximated well by the function $\vartheta_\mrm{e}(z) \propto e^{-\beta_\vartheta z/H}$ at low heights, where $\beta_\vartheta$ is independent of $\Theta$ and $L$. This allows us to write
\begin{equation}
\frac{\diff \ln \vartheta_\mrm{e}}{\diff z} = -\frac{\beta_\vartheta}{H}. \label{eq:theta_e_slope}
\end{equation}
The characteristic vertical scale $H$ is linked to the heating amplitude $\Theta$ by Eq.~\ref{eq:oom_h} and is thus relevant close to the differentially heated surface, where the typical temperature fluctuations are of the order of $\Theta$. A straightforward generalisation of Eq.~\ref{eq:theta_e_slope} is obtained by replacing $H$ by the local, height-dependent estimate $h(z)$ given by Eq.~\ref{eq:oom_h_local}. Upon doing so, we have
\begin{equation}
\frac{\diff \ln \vartheta_\mrm{e}^\prime}{\diff z^\prime} = -\frac{\beta_\vartheta}{h^\prime(z^\prime)}, \label{eq:theta_e_slope2}
\end{equation}
where we have introduced the new variables $\vartheta_\mrm{e}^\prime(z) = \vartheta_\mrm{e}(z)/\Theta$, $z^\prime = z/H,$ and $h^\prime(z) = h(z)/H$. By Eqs.~\ref{eq:oom_h} and \ref{eq:oom_h_local} we have
\begin{equation}
h^\prime(z^\prime) \approx \vartheta_\mrm{e}^\prime(z^\prime)^{1/7} \label{eq:h_prime}
,\end{equation}
and Eq.~\ref{eq:theta_e_slope2} becomes
\begin{equation}
\frac{\diff \ln \vartheta_\mrm{e}^\prime}{\diff z^\prime} = -\beta \vartheta_\mrm{e}^{\prime\,-1/7}, \label{eq:theta_e_slope3}
\end{equation}
where $\beta$ may differ slightly from $\beta_\vartheta$, because we have used an order-of-magnitude relation in the last step. Equation~\ref{eq:theta_e_slope3} shows that $\ln \vartheta_\mrm{e}^\prime(z)$ decreases with a fairly constant slope over a few orders of magnitude, but the slope starts to change as soon as a wider dynamic range is considered. Since the slope is proportional to $\vartheta_\mrm{e}^{\prime\,-1/7}$, Eq.~\ref{eq:theta_e_slope3} describes a runaway process. Indeed, the solution is
\begin{equation}
\vartheta_\mrm{e}^\prime(z^\prime) = \left[\vartheta_\mrm{e}^\prime(0)^{1/7} - \frac{\beta}{7}z^\prime\right]^7 \label{eq:theta_e}
\end{equation}
and vanishes at a finite height of $z^\prime_0 = 7/\beta$. The constant $\vartheta_\mrm{e}^\prime(0)^{1/7}$ must be very close to unity as $\vartheta_\mrm{e}^\prime(0) = \vartheta_\mrm{e}(0)/\Theta \approx 1,$ and we can simplify Eq.~\ref{eq:theta_e} to obtain
\begin{equation}
\vartheta_\mrm{e}(z) \propto \left(1 - \frac{\gamma}{7}\frac{z}{H}\right)^7, \label{eq:theta_e2}
\end{equation}
where we have returned to the non-primed variables and introduced a new constant $\gamma = \beta\vartheta_\mrm{e}^\prime(0)^{-1/7}$, which is a parameter to be adjusted to fit the numerical data. Using the local scaling relations, Eqs.~\ref{eq:oom_u_local}, \ref{eq:oom_w_local}, and \ref{eq:oom_h_local}, we derive the functional dependencies
\begin{align}
u_\mrm{e}(z) &\propto \left(1 - \frac{\gamma}{7}\frac{z}{H}\right)^4, \label{eq:u_e} \\
w_\mrm{e}(z) &\propto \left(1 - \frac{\gamma}{7}\frac{z}{H}\right)^5, \label{eq:w_e} \\
h(z) &\propto 1 - \frac{\gamma}{7}\frac{z}{H}. \label{eq:h_local}
\end{align}
The functions $\vartheta_\mrm{e}(z)$, $u_\mrm{e}(z)$, and $w_\mrm{e}(z)$ are shown in Fig.~\ref{fig:rms_tuw}. The constants of proportionality in Eqs.~\ref{eq:theta_e2}, \ref{eq:u_e}, and \ref{eq:w_e} have been adjusted independently, but all three functions share the value $\gamma = 1.3$. The good fit indicates that our line of reasoning is probably correct.

Can we conclude that the flow stops at the finite height we have just derived? No, since the scaling relations only work in the high-Re regime. Provided that Re is high close to $z = 0$, the flow speed quickly decreases according to Eqs.~\ref{eq:u_e} and \ref{eq:w_e} until $\mrm{Re} \approx 1$ is achieved at some height $z_1 < 7H/\gamma$. The weak flow above this point is supported by viscosity and gradually vanishes as $z \to \infty$.

\subsection{Allowing for a buoyancy-frequency gradient}
\label{sec:buoyancy_gradient}

So far, we have assumed that the flow occurs in a particularly simple type of thermal stratification --- one characterised by a typical buoyancy frequency $N_\mrm{typ} = \mrm{const.}$ Nevertheless, we aim to apply our results to the immediate vicinity of a convection zone, i.e. to a medium, in that the buoyancy frequency rises continuously from zero to a finite value. In this section, we first show how to estimate the  value of $N_\mrm{typ}$ in such a setting and then reapply the techniques developed in Sect.~\ref{sec:improving_the_model} to demonstrate how the varying buoyancy frequency affects the global flow field.

To do this, we have to recover the dependence of all the relevant flow properties on $N_\mrm{typ}$ by returning to a system of physical units. We recall that we use $1/N_\mrm{typ}$ as a unit of time and $(\varkappa/N_\mrm{typ})^{1/2}$ as a unit of distance, which implies that the unit of velocity is $(\varkappa N_\mrm{typ})^{1/2}$ and the unit of acceleration (hence of $\vartheta$) is $(\varkappa N_\mrm{typ}^3)^{1/2}$. We use these conversion factors throughout this section without mentioning them further. The height of the bottommost overturning cell is by Eq.~\ref{eq:h_fit}
\begin{equation}
H_\mrm{ph} = 1.3\left(\frac{\varkappa}{N_\mrm{typ}^2}\right)^{2/7} \Theta_\mrm{ph}^{1/7} L_\mrm{ph}^{2/7}, \label{eq:h_ph}
\end{equation}
where we have introduced the index ,,ph'' to indicate the use of physical units for quantities that are dimensionless in the rest of our analysis. The buoyancy frequency $N$ is now an increasing function of $z_\mrm{ph}$ and can be approximated by Eqs.~\ref{eq:N2} and \ref{eq:delta_nabla},
\begin{equation}
N(z_\mrm{ph}) = \left(\frac{\alpha g}{H_p^2}\right)^{1/2} z_\mrm{ph}^{1/2}. \label{eq:N}
\end{equation}
The overall scale of the flow pattern is given by the bottommost overturning cell, which is thus the most important. Therefore we estimate $N_\mrm{typ} = N(H_\mrm{ph}/2)$, i.e.
\begin{equation}
N_\mrm{typ} = \left(\frac{\alpha g}{H_p^2}\right)^{1/2} \left(\frac{H_\mrm{ph}}{2}\right)^{1/2}, \label{eq:N_typ}
\end{equation}
and combine Eqs.~\ref{eq:h_ph} and \ref{eq:N_typ} to obtain
\begin{equation}
H_\mrm{ph} \doteq 1.4\left(\frac{\varkappa^2 H_p^6}{\alpha^2 g}\right)^{1/9} \left(\frac{\Delta T}{T_\mrm{m}}\right)^{1/9} \left(\frac{L_\mrm{ph}}{H_p}\right)^{2/9}, \label{eq:h_ph2}
\end{equation}
where we have also expanded $\Theta_\mrm{ph} = g\Delta T/T_\mrm{m}$ to emphasise the dependence on the imposed temperature fluctuation $\Delta T/T_\mrm{m}$. We use the sign $\doteq$ in Eq.~\ref{eq:h_ph2} and also in Eqs.~\ref{eq:u_ph}, \ref{eq:w_ph}, and \ref{eq:tau_ph} below to indicate that we do not expect these estimates to be off by more than a few tens of percent. The dependence of $H_\mrm{ph}$ on the heating amplitude and length scale is somewhat weaker in Eq.~\ref{eq:h_ph2} compared with Eq.~\ref{eq:h_ph}, because Eq.~\ref{eq:h_ph2} takes into account that any gain in the flow's vertical extent brings about an increase in the typical buoyancy frequency, which in turn makes further penetration harder. This effect can also be seen when we express the characteristic velocity components and the flow's dynamical time scale in physical units,
\begin{align}
U_\mrm{ph} &\doteq 0.8\left(\frac{\varkappa g^4 H_p^3}{\alpha}\right)^{1/9} \left(\frac{\Delta T}{T_\mrm{m}}\right)^{5/9} \left(\frac{L_\mrm{ph}}{H_p}\right)^{1/9}, \label{eq:u_ph} \\
W_\mrm{ph} &\doteq 3\left(\frac{\varkappa g}{\alpha}\right)^{1/3} \left(\frac{\Delta T}{T_\mrm{m}}\right)^{2/3} \left(\frac{L_\mrm{ph}}{H_p}\right)^{-2/3}, \label{eq:w_ph} \\
\tau_\mrm{ph} &\doteq 0.7\left(\frac{\alpha H_p^6}{\varkappa g^4}\right)^{1/9} \left(\frac{\Delta T}{T_\mrm{m}}\right)^{-5/9} \left(\frac{L_\mrm{ph}}{H_p}\right)^{8/9}, \label{eq:tau_ph}
\end{align}
where the exponents have slightly changed compared with Eqs.~\ref{eq:u_fit}, \ref{eq:w_fit}, and \ref{eq:tau_fit}.

The spatial variation of $N$ brings on a first-order effect, too; that is, the stratification offers less resistance to overturning in the bottom part of the flow field compared with the rest of it. We mimic this effect by using the flow's excellent scaling properties under the assumption that the flow behaves
locally as if $N$ was constant. Our goal is to improve upon the envelope models of $\vartheta_\mrm{rms}(z)$, $u_\mrm{rms}(z),$ and $w_\mrm{rms}(z)$ derived in Sect.~\ref{sec:improving_the_model} by taking the dependence of $N$ on height into account.

Our starting point is Eq.~\ref{eq:theta_e_slope2} with the difference that now we define $\vartheta_\mrm{e}^\prime = \vartheta_\mrm{e,ph}/\Theta_\mrm{ph}$, $z^\prime = z_\mrm{ph}/H_\mrm{ph}$ and $h^\prime = h_\mrm{ph}/H_\mrm{ph}$. We caution the reader that $\vartheta_\mrm{e,ph}$ refers to a model with $N = N(z)$ and not to a direct translation of $\vartheta_\mrm{e}$ that appears in Eq.~\ref{eq:theta_e_slope2} to physical units. The local vertical length scale of the flow, $h(z)$ given by Eq.~\ref{eq:oom_h_local}, can be translated to physical units directly,
\begin{equation}
h_\mrm{ph} \approx \varkappa^{2/7} N^{-4/7} \vartheta_\mrm{e,ph}^{1/7} L_\mrm{ph}^{2/7}.
\end{equation}
This equation, together with Eq.~\ref{eq:h_ph}, implies
\begin{equation}
h^\prime \approx N^{\prime-4/7} \vartheta_\mrm{e}^{\prime 1/7},
\end{equation}
where $N^\prime = N/N_\mrm{typ} = (2 z^\prime)^{1/2}$ (see Eqs.~\ref{eq:N} and \ref{eq:N_typ}). It is evident that $h^\prime$ diverges for $N^\prime \to 0^+$, i.e. $z^\prime \to 0^+$. This effect is purely artificial because the divergence occurs within the bottommost overturning cell of the flow, and the large-scale model we are developing here cannot capture such local phenomena. We ignore the divergence for now because only $h^{\prime -1}$ appears in Eq.~\ref{eq:theta_e_slope2} and use the same procedure as in Sect.~\ref{sec:improving_the_model} to derive a generalised version of Eq.~\ref{eq:theta_e_slope3},
\begin{equation}
\frac{\diff \ln \vartheta_\mrm{e}^\prime}{\diff z^\prime} = -\beta z^{\prime 2/7} \vartheta_\mrm{e}^{\prime\,-1/7}, \label{eq:theta_e_slope4}
\end{equation}
where the parameter $\beta$ has absorbed all coefficients of the order of unity. Its value should still be of the order of unity, but it may be different in this model compared with the model developed in  Sect.~\ref{sec:improving_the_model}. By analogy to the derivation in Sect.~\ref{sec:improving_the_model}, we can write the solution to Eq.~\ref{eq:theta_e_slope4} in the form
\begin{equation}
\vartheta_\mrm{e,ph}(z_\mrm{ph}) \propto \left[1 - \frac{\gamma}{9}\left(\frac{z_\mrm{ph}}{H_\mrm{ph}}\right)^{9/7}\right]^7, \label{eq:theta_e3}
\end{equation}
where we have also returned to the non-primed quantities, and $\gamma = \beta\vartheta_\mrm{e}^\prime(0)^{-1/7}$ is a parameter of the order of unity. The typical velocity components and the typical vertical vertical length scale can be estimated using the local scaling relations, Eqs.~\ref{eq:oom_u_local}, \ref{eq:oom_w_local}, \ref{eq:oom_h_local}, and~\ref{eq:theta_e3}. We obtain
\begin{align}
u_\mrm{e,ph}(z_\mrm{ph}) &\propto \left[\frac{N(z_\mrm{ph})}{N_\mrm{typ}}\right]^{-2/7} \left[1 - \frac{\gamma}{9}\left(\frac{z_\mrm{ph}}{H_\mrm{ph}}\right)^{9/7}\right]^4, \label{eq:u_e2} \\
w_\mrm{e,ph}(z_\mrm{ph}) &\propto \left[\frac{N(z_\mrm{ph})}{N_\mrm{typ}}\right]^{-6/7} \left[1 - \frac{\gamma}{9}\left(\frac{z_\mrm{ph}}{H_\mrm{ph}}\right)^{9/7}\right]^5, \label{eq:w_e2} \\
h_\mrm{ph}(z_\mrm{ph}) &\propto \left[\frac{N(z_\mrm{ph})}{N_\mrm{typ}}\right]^{-4/7} \left[1 - \frac{\gamma}{9}\left(\frac{z_\mrm{ph}}{H_\mrm{ph}}\right)^{9/7}\right], \label{eq:h_local2}
\end{align}
where an explicit dependence on $N$ appears after the transition to physical units. These expressions diverge for $z \to 0^+$ where $N \to 0^+$ (see Eq.~\ref{eq:N}), which is just another illustration of the envelope models' inability to capture local phenomena (see also the discussion above). The bottommost part of the flow should in reality behave approximately as if it was in a medium with $N = N_\mrm{typ} = \mrm{const.}$, so we can cut off the problematic part of the $N(z)$ profile and use, for example, the function
\begin{equation}
\widetilde{N}(z) = 
\begin{cases}
N_\mrm{typ} & \text{for } 0 \leq z_\mrm{ph} \leq \frac{1}{2}H_\mrm{ph} \\
N_\mrm{typ}\left(\frac{2z_\mrm{ph}}{H_\mrm{ph}}\right)^{1/2} & \text{for } z_\mrm{ph} > \frac{1}{2}H_\mrm{ph}
\end{cases}
\label{eq:N_tilde}
\end{equation}
instead of $N(z)$ in practical calculations. Doing so makes the right-hand sides of Eqs.~\ref{eq:u_e2}, \ref{eq:w_e2}, and \ref{eq:h_local2} converge to unity as $z_\mrm{ph} \to 0^+$.

Just as the results of Sect.~\ref{sec:improving_the_model} do not mean that the flow vanishes at a finite height, neither the results of this section mean that. Again, the sudden drop in the typical velocities predicted by Eqs.~\ref{eq:u_e2} and \ref{eq:w_e2} only signifies that the flow undergoes a transition to the low-Re regime at a relatively low height. Eqs.~\ref{eq:u_e2} and \ref{eq:w_e2} cease to be usable from that point on and the weak flow supported by viscosity gradually vanishes as $z_\mrm{ph} \to \infty$.

\section{Application to stellar conditions}
\label{sec:application_to_stars}

The flow in a layer of thickness $h_\mrm{ph}(z_\mrm{ph})$ and vertical velocity $w_\mrm{e,ph}(z_\mrm{ph})$  at distance $z_\mrm{ph}$ from the boundary overturns a passive tracer in it on a time scale $\tau_{\rm m}=h_\mrm{ph}/w_\mrm{e,ph}$. This suggests an effective diffusion coefficient $D_{\rm eff}\approx h_\mrm{ph}w_\mrm{e,ph}$. For the first layer above the boundary, this is
\begin{equation}
D_{\rm eff}(0) = W_\mrm{ph} H_\mrm{ph}. \label{eq:D0}
\end{equation}
At distance $z_\mrm{ph}$, Eqs.~\ref{eq:w_e2} and \ref{eq:h_local2} give
\begin{equation}
D_{\rm eff}(z_\mrm{ph}) = D_{\rm eff}(0) \left[\frac{\widetilde{N}(z_\mrm{ph})}{N_\mrm{typ}}\right]^{-10/7} \left[1 - \frac{\gamma}{9}\left(\frac{z_\mrm{ph}}{H_\mrm{ph}}\right)^{9/7}\right]^6, \label{eq:D}
\end{equation}
where $N_\mrm{typ}$ is given by Eq.~\ref{eq:N_typ} and we have replaced $N(z_\mrm{ph})$ in Eqs.~\ref{eq:w_e2} and \ref{eq:h_local2} by $\widetilde{N}(z_\mrm{ph})$ given by Eq.~\ref{eq:N_tilde} as discussed in Sect.~\ref{sec:buoyancy_gradient}. The constant $\gamma$ is of the order of unity but cannot be constrained further by our present analysis. It determines the maximum height $z_\mrm{max,ph}$ that the mixing process can reach, $z_\mrm{max,ph} = (9/\gamma)^{7/9}H_\mrm{ph}$. 

For a specific example, consider the boundary of the core convection zone in a $10\,M_\odot$ zero age main sequence star. This environment is characterised by $\alpha = \diff (\nabla_\mrm{ad} - \nabla) / \diff (z_\mrm{ph}/H_p) = 0.14$, a thermal diffusivity $\varkappa = 5.9\times 10^{10}$\,cm$^2$\,s$^{-1}$, a gravitational acceleration $g = 1.1\times 10^5$\,cm\,s$^{-2}$ and a pressure scale height $H_\mrm{p} = 2.9\times 10^{10}$\,cm. A mixing-length estimate for convection in the core produces temperature fluctuations $\Delta T/T_\mrm{m} \approx 10^{-6}$ on a horizontal length scale $L_\mrm{ph} \approx H_\mrm{p}$. Equation~\ref{eq:h_ph2} then predicts that the typical vertical length scale is \mbox{$H_\mrm{ph} \approx 2\times 10^8\,\mrm{cm} =  7\times 10^{-3} H_\mrm{p}$}. The typical vertical velocity (Eq.~\ref{eq:w_ph}) is $W_\mrm{ph} \approx 5\times 10^1\,\mrm{cm}\,\mrm{s}^{-1}$. These numbers imply \mbox{$\mrm{Pe}_z = (W_\mrm{ph} H_\mrm{ph})/\varkappa \approx 2$}; i.e., the bottom part of the flow is located right at the transition between the regions of advection-dominated and diffusion-dominated heat transport. This is not a coincidence, because we are modelling the region where heat leaks from the convective eddies, allowing them to turn over and sink back to the convection zone. Such a flow has to have $\mrm{Pe}_z \approx 1$. Therefore, the effective diffusivity close to the convection zone, $D_{\rm eff}(0)$ in Eq.~\ref{eq:D}, is of the same order as the diffusivity of heat $\varkappa$. Diffusivities that are several orders of magnitude smaller than $\varkappa$ can be important on the long nuclear time scale. The maximum height reached by the differential heating process on this time scale can thus be approximated by $z_\mrm{max,ph}$. Assuming $\gamma = 1$ we obtain $z_\mrm{max,ph} \approx 4\times 10^{-2} H_\mrm{p}$.

Equation~(\ref{eq:D}) is likely to be somewhat of an overestimate of the actual mixing rate of the differential-heating process. The layers mix on the hydrodynamic time scale in their interiors, but as long as they are stationary, transport of the tracer between layers takes place by diffusion. As in the case of semiconvective layering \citep[cf.][]{Spruit13}, this reduces the effective mixing rate to the geometric mean of the microscopic diffusion coefficient $\kappa_\mrm{t}$ of the tracer and the estimate (\ref{eq:D}).

More significantly, the picture is complicated by the time dependence of the convective heat source. For the $10 M_\odot$ example, only the bottommost part of the flow can approach the stationary flow speed before the heating pattern changes because the dynamical time scale $\tau_\mrm{ph} \approx 5\times 10^6\,\mrm{s}$ (Eq.~\ref{eq:tau_ph} with the parameter values stated above) is of the same order as the convective overturning time scale in the core. This is likely to lead to some form of averaging, reducing the effective amplitude of the source. The level of this effect can probably be investigated with a time-dependent simulation.

\section{Summary}

Various observations show that there is a need for some additional mixing at the interfaces between the convective and radiative layers of stars. Even processes that are too weak to be detectable in numerical hydrodynamic simulations need to be considered as candidate sources of this mixing, because the nuclear time scale on the main sequence is so much longer than the dynamical time scale of convection, and cumulative effects are likely to play an important role. 

In this work, we have investigated one such weak process, which we call ``differential heating''. The differential heating process occurs when radiative diffusion transports a temperature fluctuation from the boundary of a convection zone into the neighbouring stable stratification. The resulting perturbation of hydrostatic equilibrium triggers a weak flow, which may provide mixing up to some distance from the convection zone. We investigated  the flow that is driven by a static temperature fluctuation varying sinusoidally along the solid horizontal boundary of a stably stratified, thin layer of gas. This low-P\'eclet number problem (i.e. a slow flow dominated by thermal diffusion) turns out to be intrinsically nonlinear, in the sense that the horizontal structure of the flow is asymmetric. Even for symmetric boundary conditions, the upflow is narrower than the downflowing part for the flow, and the shape of the flow pattern is nearly independent of the amplitude of the driving temperature perturbation.

A few additional assumptions (Sect.~\ref{sec:differential_heating_problem}) allow us to describe the problem by a set of dimensionless equations, the solution to which depends (apart from the boundary and initial conditions) only on the Prandtl number. We analysed these differential-heating equations for their scaling properties under the assumption that the flow is stationary (Sect.~\ref{sec:analytical_considerations}). An astrophysically interesting corner of the parameter space is characterised by $\mrm{Re}_x \gg 1$, $\mrm{Re}_z \gg 1$, $\mrm{Pe}_x \gg \mrm{Pe}_z$, and $\mrm{Pe}_z \ll 1.$ (The $x$ and $z$ directions have to be distinguished because such flow has a high aspect ratio.) In this limit we derive a set of simple relations (Eqs.~\ref{eq:oom_h} and \ref{eq:oom_u}\,--\,\ref{eq:oom_tau}) to describe how the global flow properties depend on the heating amplitude $\Theta$ and length scale $L$. We find, in particular, that the characteristic vertical length scale $H$ depends only weakly on the heating parameters (Eq.~\ref{eq:oom_h}).

We developed a dedicated numerical code to solve the equations. The main difficulties are related to the highly diffusive nature of the flow, its high aspect ratio, and the need to resolve a wide dynamic range in the flow amplitude within the computational box (as much as five orders of magnitude). The flow in our two-dimensional, time-dependent simulations reaches a stationary state at all values of the Reynolds number that we have been able to achieve (up to $\mrm{Re} \equiv \mrm{Re}_x = \mrm{Re}_z = 4\times 10^3$). The flow is always composed of several layers of overturning cells, the shape of which depends only on the Reynolds number and not on the heating length scale $L$ and amplitude $\Theta$. This property makes the flow scaleable in the sense that the flow field corresponding to some heating parameters $L_1$, $\Theta_1$ can be stretched in space and scaled in amplitude to get a good approximation of the flow field corresponding to a different set of heating parameters $L_2$, $\Theta_2$ provided that $\mrm{Re}$ is in both cases the same. This is also the reason the scaling relations derived in Eq.~\ref{sec:analytical_considerations} fit the simulation data remarkably well at $\mrm{Re} = \mrm{const.}$ (see Fig.~\ref{fig:scaling_relations}). Increasing the Reynolds number has little influence on the flow speed, but it makes the flow pattern increasingly asymmetric. 

We decrease the artificial-viscosity coefficients in the code with height in order to keep the Reynolds number approximately the same in every layer of flow cells. The numerical data show that the global scaling relations derived in Sect.~\ref{sec:analytical_considerations} have their local analogues, which can be used within the flow. The flow speed's decrease with height, being locally exponential, steepens with the decreasing flow amplitude according to the local scaling relations. Based on this we derive a model of the flow's dependence on height, which closely fits the numerical data over the whole dynamic range that we have been able to cover (as much as five orders of magnitude, see Fig.~\ref{fig:rms_tuw}). The model shows that the flow speed drops abruptly to a negligible value at a finite height. The local scaling relations also allow us to generalise our results to the more realistic case, in which the buoyancy frequency $N$ increases with height (see Sect.~\ref{sec:improving_the_model}).

We illustrated the typical scales associated with the stellar differential-heating process with the example of the convective core of a $10\,M_\odot$ zero-age main sequence star (see Sect.~\ref{sec:application_to_stars}). We approximate the mixing due to the differential-heating flow by an ``effective'' diffusion coefficient $D_\mrm{eff}$, which is of the order of the diffusivity of heat near the convection zone and decreases with height according to Eq.~\ref{eq:D}. The mixing relevant for stellar evolution extends about $4\%$ of the pressure scale height above the convection zone.

\subsection{Main findings of the paper}

\begin{enumerate}[label={(\arabic*)}]
\item The flow has a cellular structure and reaches a stationary state at all values of the Reynolds number that we have been able to achieve (up to $\mrm{Re} = 4\times 10^3$).
\item Both global and local properties of the flow can be described by a set of simple analytical relations.
\item The flow speed drops abruptly to a negligible value at a finite height above the source of heating.
\item The mixing relevant for stellar evolution extends about $4\%$ of the pressure scale height above the convection zone of a $10\,M_\odot$ zero-age main sequence star.
\end{enumerate}

\section*{Acknowledgements}

We would like to thank Achim Weiss for the $10M_\odot$ model, Ewald M\"uller and Maxime Viallet for enlightening discussions on numerical hydrodynamics, and the anonymous referee for critical comments that improved the overall presentation of the text.

\appendix
\clearpage
\onecolumn
\section{Numerical methods}
\label{sec:numerical_methods}

\subsection{Integration scheme}
\label{sec:integration_scheme}

We have adapted the standard MacCormack method to suit our specific problem. In the simplest case of a one-dimensional vector $\bm{q}$ of conserved quantities being advected on and equidistant grid with a spacing of $\Delta x$, MacCormack's method can be written as
\begin{align}
\bm{q}_k^{(1)} &= \bm{q}_k^n - \Delta t \frac{\bm{f}\left(\bm{q}_{k+1}^n\right) - \bm{f}\left(\bm{q}_{k}^{n\phantom{|}}\right)}{\Delta x},\label{eq:MacCormack1} \\
\bm{q}_k^{(2)} &= \bm{q}_k^{(1)} - \Delta t \frac{\bm{f}\left(\bm{q}_k^{(1)}\right) - \bm{f}\left(\bm{q}_{k-1}^{(1)}\right)}{\Delta x},\label{eq:MacCormack2} \\
\bm{q}_k^{n+1} &= \frac{\bm{q}_k^n + \bm{q}_k^{(2)}}{2},\label{eq:MacCormack3}
\end{align}
where $\bm q_k^n$ is the value of $\bm q$ at the $k$-th grid point and the $n$-th time step, $\Delta t$ the time step, $\bm{f}(\bm{q})$ the flux function, and we use the convention that any parenthesised upper index refers to a sub-step of the method instead of a time-step index. The method is linearly stable provided that the CFL condition $\Delta t \leq \Delta x / \rho(A)$ is met, where $A$ is the Jacobian matrix of the flux vector and $\rho(A)$ is the largest characteristic value of $A$. Nonlinear stability typically requires the addition of some form of artificial viscosity. MacCormack's method is second-order accurate both in space and time.

We discretise Eqs.~\ref{eq:continuity2}, \ref{eq:energy2}, and \ref{eq:momentum4} on a collocated, two-dimensional grid of $M \times N$ cells with constant cell spacing $(\Delta x,\, \Delta z)$. The two spatial dimensions and the presence of source terms in the equations forces us to significantly extend the basic MacCormack scheme. We begin by advecting the vector of variables $\bm{q} = (u,\, w,\, \vartheta)$ in both spatial directions using Strang splitting,
\begin{align}
\bm{q}_{k,l}^{(1\mrm{a})} &= \bm{q}_{k,l}^n - \frac{\Delta t}{2}\frac{u_{k+1,l}^n\, \bm{q}_{k+1,l}^n - u_{k,l}^n\, \bm{q}_{k,l}^n}{\Delta x},\\
\bm{q}_{k,l}^{(1\mrm{b})} &= \bm{q}_{k,l}^{(1\mrm{a})} - \Delta t\frac{w_{k,l+1}^{(1\mrm{a})}\, \bm{q}_{k,l+1}^{(1\mrm{a})} - w_{k,l}^{(1\mrm{a})}\, \bm{q}_{k,l}^{(1\mrm{a})}}{\Delta z},\\
\bm{q}_{k,l}^{(1\mrm{c})} &= \bm{q}_{k,l}^{(1\mrm{b})} - \frac{\Delta t}{2}\frac{u_{k+1,l}^{(1\mrm{b})}\, \bm{q}_{k+1,l}^{(1\mrm{b})} - u_{k,l}^{(1\mrm{b})}\, \bm{q}_{k,l}^{(1\mrm{b})}}{\Delta x},
\end{align}
where we have written out the explicit form of the flux terms. The indices $k$ and $l$ refer to the position along the $x$ and $z$ axes, respectively. We proceed by adding the source terms to the momentum equations,
\begin{align}
u_{k,l}^{(1\mrm{d})} &= u_{k,l}^{(1\mrm{c})} + \Delta t\left[-\frac{p_{k+1,l}^n - p_{k-1,l}^n}{2\Delta x} + \nu_l\frac{u_{k-1,l}^{(1\mrm{c})} - 2u_{k,l}^{(1\mrm{c})} + u_{k+1,l}^{(1\mrm{c})}}{(\Delta x)^2} + \frac{\mu_{l+1/2}\left(u_{k,l+1}^{(1\mrm{c})} - u_{k,l}^{(1\mrm{c})}\right) - \mu_{l-1/2}\left(u_{k,l}^{(1\mrm{c})} - u_{k,l-1}^{(1\mrm{c})}\right)}{(\Delta z)^2}\right], \\
w_{k,l}^{(1\mrm{d})} &= w_{k,l}^{(1\mrm{c})} + \Delta t\left[-\frac{p_{k,l+1}^n - p_{k,l-1}^n}{2\Delta z} + \vartheta_{k,l}^{(1\mrm{c})} + \nu_l\frac{w_{k-1,l}^{(1\mrm{c})} - 2w_{k,l}^{(1\mrm{c})} + w_{k+1,l}^{(1\mrm{c})}}{(\Delta x)^2} + \frac{\mu_{l+1/2}\left(w_{k,l+1}^{(1\mrm{c})} - w_{k,l}^{(1\mrm{c})}\right) - \mu_{l-1/2}\left(w_{k,l}^{(1\mrm{c})} - w_{k,l-1}^{(1\mrm{c})}\right)}{(\Delta z)^2}\right],
\end{align}
where we use second-order-accurate central differences to keep up with the order of accuracy of the advection scheme, $\nu_l = \mrm{Pr}_x(z_l)$ and $\mu_l = \mrm{Pr}_z(z_l)$ are the coefficients of our anisotropic artificial-viscosity prescription (see Sect.~\ref{sec:numerical_solutions}), and $\mu_{l+1/2} = (\mu_l + \mu_{l+1})/2$. The new velocity field $\bm{u}^{(1\mrm{d})} = \left(u^{(1\mrm{d})},\, w^{(1\mrm{d})}\right)$ is, in general, slightly divergent. We correct for this divergence by subtracting the gradient of a pressure-correction field, $\bm{u}^{(1)} = \bm{u}^{(1\mrm{d})} - \Delta t\, \bm{\nabla} (\Delta p)^{(1)}$. The condition \mbox{$\bm{\nabla}\cdot \bm{u}^{(1)} = 0$} leads to a Poisson equation for the pressure correction,
\begin{align}
\bm\nabla^2(\Delta p)^{(1)} = \frac{\bm\nabla\cdot \bm u^{(1\mrm{d})}}{\Delta t}.
\label{eq:pressure_correction}
\end{align}
Since we use central differences to compute partial derivatives, the discrete form of the Laplace operator in Eq.~\ref{eq:pressure_correction} should be derived by applying the central differences twice. That would, however, lead to a sparse operator and cause odd-even-decoupling problems on our collocated grid. Therefore we use the standard compact Laplacian and solve the approximate pressure-correction equation
\begin{align}
\frac{(\Delta p)_{k-1,l}^{(1)} - 2(\Delta p)_{k,l}^{(1)} + (\Delta p)_{k+1,l}^{(1)}}{(\Delta x)^2} + \frac{(\Delta p)_{k,l-1}^{(1)} - 2(\Delta p)_{k,l}^{(1)} + (\Delta p)_{k,l+1}^{(1)}}{(\Delta z)^2} = \frac{1}{\Delta t} \left[\frac{u_{k+1,l}^{(1\mrm{d})} - u_{k-1,l}^{(1\mrm{d})}}{2\Delta x} + \frac{w_{k,l+1}^{(1\mrm{d})} - w_{k,l-1}^{(1\mrm{d})}}{2\Delta z}\right].
\label{eq:pressure_correction2}
\end{align}
Equation~\ref{eq:pressure_correction2} is solved by a spectral solver, see Sect.~\ref{sec:spectral_solvers}. Having computed the pressure correction, we apply it to the velocity field,
\begin{align}
u_{k,l}^{(1)} &= u_{k,l}^{(1\mrm{d})} - \Delta t \frac{(\Delta p)_{k+1,l}^{(1)} - (\Delta p)_{k-1,l}^{(1)}}{2\Delta x}, \\
w_{k,l}^{(1)} &= w_{k,l}^{(1\mrm{d})} - \Delta t \frac{(\Delta p)_{k,l+1}^{(1)} - (\Delta p)_{k,l-1}^{(1)}}{2\Delta z}.
\end{align}
The approximate nature of the pressure-correction equation (Eq.~\ref{eq:pressure_correction2}) causes $\bm{\nabla}\cdot \bm{u}^{(1)}$ to be small, but non-zero. Practical experience has shown that the residual divergence is negligibly small in the flows analysed in this paper provided that the boundary conditions are treated properly, see Sect.~\ref{sec:boundary_conditions}. We should also write $p_{k,l}^{(1)} = p_{k,l}^{n} + (\Delta p)_{k,l}^{(1)}$ at this point, but our numerical tests have shown that the residual divergence in the velocity field becomes much smaller if we set $p_{k,l}^{(1)} = p_{k,l}^{n}$, so we use the latter form. The next step is to integrate the remaining two terms in the energy equation. We begin by adding the $-w$ term,
\begin{equation}
\vartheta_{k,l}^{(1\mrm{d})} = \vartheta_{k,l}^{(1\mrm{c})} - \Delta t\, w_{k,l}^{(1)},
\end{equation}
where its latest available value, $-w^{(1)}$, has been used. The diffusion sub-step is given by the implicit equation
\begin{equation}
\vartheta_{k,l}^{(1)} = \vartheta_{k,l}^{(1\mrm{d})} + \Delta t \left[\frac{\vartheta_{k-1,l}^{(1)} - 2\vartheta_{k,l}^{(1)} + \vartheta_{k+1,l}^{(1)}}{(\Delta x)^2} + \frac{\vartheta_{k,l-1}^{(1)} - 2\vartheta_{k,l}^{(1)} + \vartheta_{k,l+1}^{(1)}}{(\Delta z)^2}\right], \label{eq:theta_implicit}
\end{equation}
which is also solved by a spectral solver, see Sect.~\ref{sec:spectral_solvers}. We have thus completed the first step of the MacCormack scheme, analogous to Eq.~\ref{eq:MacCormack1}, and obtained the new variables $u^{(1)}$, $w^{(1)}$, $p^{(1)}$, and $\vartheta^{(1)}$. The second step, which we do not do not go into detail on, differs from the first one at two points. First, advection is done using backward-space flux differencing, as in Eq.~\ref{eq:MacCormack2} (compare with Eq.~\ref{eq:MacCormack1}). Second, the pressure field is updated in this step, i.e. $p_{k,l}^{(2)} = p_{k,l}^{(1)} + (\Delta p)_{k,l}^{(2)}$. The final step of the MacCormack's scheme, Eq.~\ref{eq:MacCormack3}, is used in the same form, with $\bm{q} = (u,\, w,\, \vartheta)$. We also update the pressure field in the same way, $p_{k,l}^{(n+1)} = \frac{1}{2}\left(p_{k,l}^n + p_{k,l}^{(2)}\right)$, so that we obtain an estimate of the pressure field for the next time step.

Finally, there is a simple way of increasing the accuracy of the scheme at a given grid resolution, which we use. The MacCormack method contains a built-in asymmetry: Eqs.~\ref{eq:MacCormack1} and \ref{eq:MacCormack2} show that it always starts with forward-space flux differencing and continues with backward-space flux differencing. The two flux-differencing methods can be reversed, obtaining a ``reverse'' MacCormack method, without decreasing the order of accuracy of the overall scheme. We compute every time step using both the ``direct'' and the ``reverse'' methods and use the arithmetic average of the estimates given by the two methods.

\subsection{Boundary conditions}
\label{sec:boundary_conditions}

\begin{figure*}
\centering
\includegraphics[width=17cm]{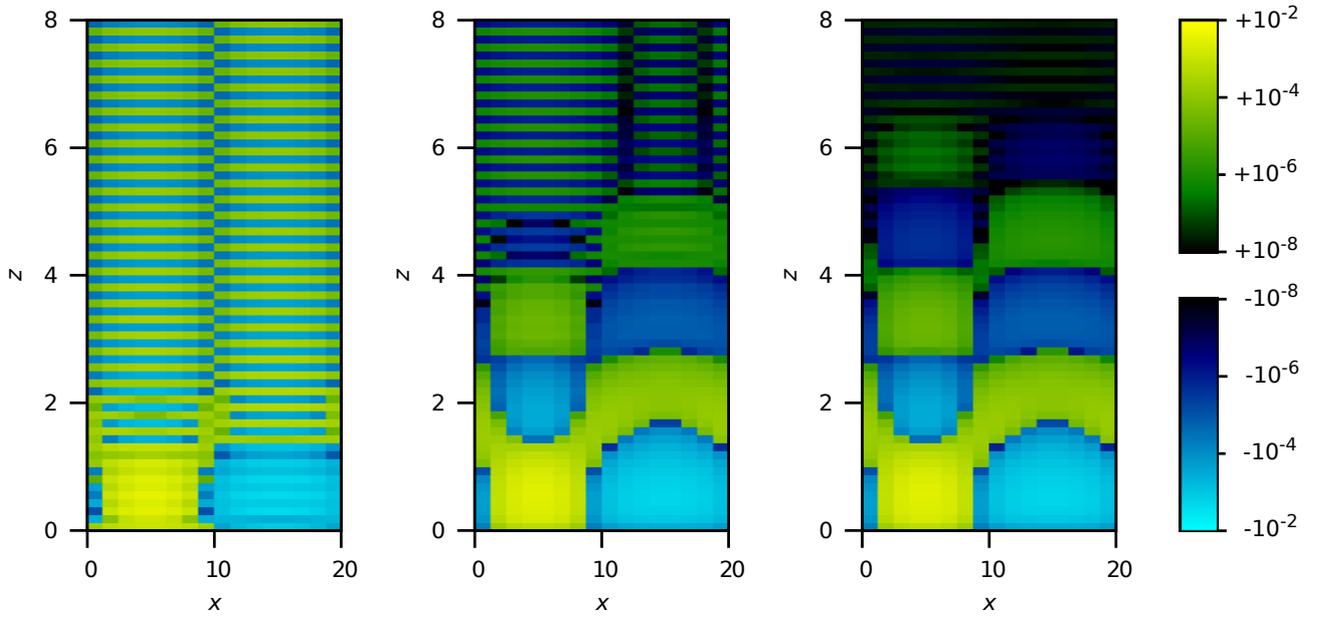}
\caption{Effect of three different methods of treating the pressure at the solid top and bottom boundaries. In all three panels, the vertical velocity component, $w$, is plotted on a split logarithmic colour scale. We use $\Theta = 10^{-3}$, $L = 10^1$, constant kinematic viscosity and set the resolution to only $16 \times 64$ to make the spurious oscillations visible. We obtain the result plotted in the left panel using the simple symmetry conditions for pressure (Eqs.~\ref{eq:bc_p1} and \ref{eq:bc_p2}). Preceding the pressure-gradient computation by third-order pressure extrapolation to the ghost cells reduces the oscillations' amplitude by a factor of $\sim 100$ (middle panel). Increasing the extrapolation order to six brings about another decrease by a factor of $\sim 30$ in the oscillations' amplitude (right panel). The pressure gradient is in all three cases computed by the second-order central differences in the whole computational domain.}
\label{fig:p_boundary_treatment}
\end{figure*}

The treatment of boundaries is restricted by our decision to use spectral solvers, which do not allow  changing the differentiation operators anywhere in the computation domain. We use the ghost-cell technique for this reason. The boundary conditions we impose on the differential-heating flow are summarised in Sect.~\ref{sec:numerical_solutions}. The periodic boundaries in the horizontal direction are trivial to implement. The solid boundaries on the top and bottom of the computational domain, however, require much more care. We implement them using reflective boundary conditions for the velocity vector,
\begin{align}
u_{k,-1} &= u_{k,0}, \label{eq:bc_u1} \\
u_{k,N} &= u_{k,N-1}, \label{eq:bc_u2} \\
w_{k,-1} &= -w_{k,0}, \label{eq:bc_w1}\\
w_{k,N} &= -w_{k,N-1}, \label{eq:bc_w2}
\end{align}
so that the imaginary walls are located at $l = -1/2$ and at $l = N - 1/2$. The conditions imposed on $u$ also eliminate any shear on the boundary. The pressure field is required to be symmetric with respect to the solid boundaries,
\begin{align}
p_{k,-1} &= p_{k,0}, \label{eq:bc_p1}\\
p_{k,N} &= p_{k,N-1}. \label{eq:bc_p2}
\end{align}
The conditions imposed by Eqs.~\ref{eq:bc_u1}--\ref{eq:bc_p2} can easily be shown to be consistent with the pressure-correction equation (Eq.~\ref{eq:pressure_correction2}; sum both sides over $k=0,\, 1,\, \ldots,\, M$ and $l=0,\, 1,\, \ldots,\, N$). They typically do, however, bring about a cusp in the pressure field along the normal to the walls. The resulting discontinuity in the vertical pressure gradient then propagates to the rest of the domain and can be seen as a low-amplitude oscillatory field superimposed on the true pressure field (see the left panel of Fig.~\ref{fig:p_boundary_treatment}). We tried to cure this problem by changing the discretisation of the vertical-gradient operator at the walls, so that the ghost cells would not be used when computing the pressure gradient. This solution has met with very little success, most likely because the abrupt change in the operator brings about an abrupt change in the discretisation error so the problem remains. Quite surprisingly, preceding the pressure-gradient computation by high-order pressure extrapolation to the ghost cells has turned out to be an effective solution, able to eliminate nearly all of the spurious oscillations (see the middle and right panels of Fig.~\ref{fig:p_boundary_treatment}). We therefore use sixth-order extrapolation in the simulations with constant artificial viscosity and increase the extrapolation order to ten when we let the artificial viscosity decrease with height. This technique cannot be viewed, however, as an all-purpose solution, because it is likely to be too unstable to be useful when computing highly turbulent flows.

We require the temperature fluctuation $\vartheta$ to have a fixed sinusoidal profile at the bottom boundary and to vanish at the upper boundary, which translates into
\begin{align}
\vartheta_{k,-1} &= -\vartheta_{k,0} + 2\Theta\sin\left(\frac{\pi x_k}{L}\right), \label{eq:bc_theta1}\\
\vartheta_{k,N} &= -\vartheta_{k,N-1}. \label{eq:bc_theta2}
\end{align}

\subsection{Spectral solvers}
\label{sec:spectral_solvers}

We use spectral methods to solve the two equations involving the Laplace operator, the Poisson equation for the pressure-correction equation (Eq.~\ref{eq:pressure_correction2}) and the implicit heat-diffusion equation (Eq.~\ref{eq:theta_implicit}). We express both the knowns and unknowns as linear combinations of the Laplacian's eigenfunctions that comply with the desired boundary conditions. The solution procedure is then much simplified and effective, provided that the transform to the eigenfunction basis can be computed efficiently.

In case of the pressure-correction equation (Eq.~\ref{eq:pressure_correction2}), we use the linear transform
\begin{equation}
\hat{f}_{m,n} = \frac{1}{2MN}\sum_{k=0}^{M-1} \left[2\sum_{l=0}^{N-1} f_{k,l} \cos\left(\frac{\pi\, n \left(l + \frac{1}{2}\right)}{N}\right)\right] \exp\left(-\frac{2\pi i m k}{M}\right) \label{eq:transform1}
\end{equation}
and its inverse
\begin{equation}
f_{k,l} = \sum_{m=0}^{M-1} \left[\hat{f}_{m,0} + 2\sum_{n=1}^{N-1} \hat{f}_{m,n} \cos\left(\frac{\pi\, n \left(l + \frac{1}{2}\right)}{N}\right)\right] \exp\left(\frac{2\pi i m k}{M}\right) \label{eq:transform2}
\end{equation}
to transform any field $f_{k,l}$ to an array of complex amplitudes $\hat{f}_{m,n}$ and back. We can see that the basis functions in Eq.~\ref{eq:transform2} are periodic in $k$ and even around $l = -1/2$ and $l = N - 1/2$; i.e., they comply with our boundary conditions on the pressure field (see Sect.~\ref{sec:boundary_conditions}). Upon using the spectral decomposition defined by Eq.~\ref{eq:transform2} on both sides of the pressure-correction equation (Eq.~\ref{eq:pressure_correction2}), we readily obtain its solution in the wavenumber space,
\begin{equation}
(\Delta \hat{p})_{m,n} = \frac{\hat{S}_{m,n}}{\lambda_{m,n}},
\end{equation}
where we have omitted the upper indices because the expression applies to both steps of the MacCormack scheme, $\hat{S}_{k,l}$ is the transformed right-hand side of Eq.~\ref{eq:pressure_correction2}. The eigenvalues $\lambda_{m,n}$ of the Laplacian are
\begin{equation}
\lambda_{m,n} = -\frac{2 - 2\cos\left(\frac{2\pi\, m}{M}\right)}{(\Delta x)^2} -\frac{2 - 2\cos\left(\frac{\pi\, n}{N}\right)}{(\Delta z)^2}
\end{equation}
and can be pre-computed. We set $\lambda_{0,0}$ to a large number to prevent division by zero and make the undetermined component $(\Delta \hat{p})_{0,0}$ vanish.

In case of the heat-diffusion equation (Eq.~\ref{eq:theta_implicit}), we use the linear transform
\begin{equation}
\hat{g}_{m,n} = \frac{1}{2MN}\sum_{k=0}^{M-1} \left[2\sum_{l=0}^{N-1} g_{k,l} \sin\left(\frac{\pi\, \left(n + 1\right) \left(l + \frac{1}{2}\right)}{N}\right)\right] \exp\left(-\frac{2\pi i m k}{M}\right) \label{eq:transform3}
\end{equation}
and its inverse
\begin{equation}
g_{k,l} = \sum_{m=0}^{M-1} \left[(-1)^l\, \hat{g}_{m,N-1} + 2\sum_{n=0}^{N-2} \hat{g}_{m,n} \sin\left(\frac{\pi\, \left(n + 1\right) \left(l + \frac{1}{2}\right)}{N}\right)\right] \exp\left(\frac{2\pi i m k}{M}\right) \label{eq:transform4}
\end{equation}
to transform any field $g_{k,l}$ to an array of complex amplitudes $\hat{g}_{m,n}$ and back. We can see that the basis functions in Eq.~\ref{eq:transform4} are periodic in $k$ and odd around $l = -1/2$ and $l = N - 1/2$; i.e., they comply with our boundary conditions on the temperature field in case of a vanishing heating amplitude (see Sect.~\ref{sec:boundary_conditions}). To allow for an arbitrary heating profile at the bottom boundary, we take out the known boundary term from the Laplacian on the right-hand side of Eq.~\ref{eq:theta_implicit} and treat it as a source term. One can show that it is the same as replacing the diffusion equation $\partial\vartheta / \partial t = \bm{\nabla}^2 \vartheta$ by the equivalent equation $\partial(\vartheta - \zeta) / \partial t = \bm{\nabla}^2 (\vartheta - \zeta)$, where $\zeta$ is the static solution to the diffusion equation $\partial\zeta / \partial t = \bm{\nabla}^2 \zeta$ with the desired boundary conditions ($\zeta$ can be pre-computed for a fixed heating profile). The boundary conditions on the difference $\vartheta - \zeta$ are then identically zero, and the spectral decomposition defined by Eq.~\ref{eq:transform4} can be used. This way we obtain an explicit expression for the solution of the implicit Eq.~\ref{eq:theta_implicit} in the wavenumber space,
\begin{equation}
\hat{\vartheta}_{m,n}^{(1)} = \frac{\hat{\vartheta}_{m,n}^{(1\mrm{d})} - \hat{\zeta}_{m,n}}{1 - \Delta t \Lambda_{m,n}} + \hat{\zeta}_{m,n} \label{eq:theta_solution}
,\end{equation}
where the eigenvalues $\Lambda_{m,n}$ of the Laplacian are
\begin{equation}
\Lambda_{m,n} = -\frac{2 - 2\cos\left(\frac{2\pi\, m}{M}\right)}{(\Delta x)^2} -\frac{2 - 2\cos\left(\frac{\pi\, (n + 1)}{N}\right)}{(\Delta z)^2}
\end{equation}
and can be pre-computed. An equation analogous to Eq.~\ref{eq:theta_solution} relates $\hat{\vartheta}^{(2)}$ to $\hat{\vartheta}^{(2\mrm{d})}$.

In the practical implementation, we use the FFTW library \citep{FrigoJohnson05} to compute the transforms in Eqs.~\ref{eq:transform1}, \ref{eq:transform2}, \ref{eq:transform3}, and \ref{eq:transform4}. We combine standard, one-dimensional transforms of different kinds to obtain the non-standard, two-dimensional transforms that we need. Namely, Eq.~\ref{eq:transform1} is implemented as a series of DCT-II transforms over the rows of the input array, after which the columns of the resulting array are transformed by a series of DTF transforms. The backward transform (Eq.~\ref{eq:transform2}) is then computed by a series of DFTs followed by a series of DCT-IIIs. The transforms for the diffusion equation (Eq.~\ref{eq:transform3} and \ref{eq:transform4}) are implemented in the same way, but simply replacing the DCT-IIs by DST-IIs and DCT-IIIs by DST-IIIs. The transforms from the FFTW library do not include the normalisation factor $(2MN)^{-1}$.

\end{document}